\documentclass[10pt]{article}
\usepackage[utf8]{inputenc}
\usepackage{amsmath}
\usepackage{amssymb}
\usepackage{graphicx}
\usepackage[table]{xcolor}
\usepackage{orcidlink}
\usepackage{hyperref}
\usepackage{graphicx}
\usepackage{epstopdf}
\usepackage{amsmath}
\usepackage{enumitem}
\usepackage{hyperref}
\usepackage{epsfig}
\usepackage{longtable}
\hypersetup{colorlinks=true, linkcolor=blue, citecolor=blue, urlcolor=blue}
\usepackage{caption}
\usepackage{subfigure}
\usepackage{subfigure}
\usepackage{booktabs} 
\usepackage{caption}  
\usepackage{geometry} 
\RequirePackage[numbers,sort&compress]{natbib}
\begin{document}
\baselineskip=16pt

\newcommand{\apjl}{APJL}
\newcommand{\apj}{APJ}
\newcommand{\aap}{AAP}
\newcommand{\mnras}{MNRAS}
\newcommand{\nar}{NAR}
\newcommand{\apjs}{APJS}
\newcommand{\jcap}{JCAP}
\newcommand{\prd}{PRD}
\newcommand{\apss}{APSS}
\newcommand{\prl}{PRL}
\newcommand{\nat}{NAT}
\newcommand{\araa}{ARAA}
\newcommand{\pasj}{PASJ}
\newcommand{\plb}{PLB}

\begin{center}
\LARGE{Swampland Conjectures through ACT Observations: Observational Signatures of Radiative-Corrected Inflation}
\end{center}

\vspace{0.3cm}
\begin{center}
{\bf Mohammad Ali S Afshar\orcidlink{0009-0001-3133-5992}}\footnote{\bf m.a.s.afshar@gmail.com}\\
{\it Department of Physics, Faculty of Basic
Sciences, University of Mazandaran\\ P. O. Box 47416-95447, Babolsar, Iran}\\
{\it School of Physics, Damghan University, Damghan 3671645667, Iran}\\
{\it Canadian Quantum Research Center, 204-3002 32 Ave Vernon, BC V1T 2L7, Canada}\\
{\bf Saeed Noori Gashti\orcidlink{0000-0001-7844-2640}}\footnote{\bf saeed.noorigashti70@gmail.com, sn.gashti@du.ac.ir}\\
{\it School of Physics, Damghan University, Damghan 3671645667, Iran}\\
{\bf Mohammad Reza Alipour\orcidlink{x}}\footnote{\bf mohamad.alipour.1994@gmail.com, mr.alipour@stu.umz.ac.ir}\\
{\it Department of Physics, Faculty of Basic
Sciences, University of Mazandaran\\ P. O. Box 47416-95447, Babolsar, Iran}\\
{\it School of Physics, Damghan University, Damghan 3671645667, Iran}\\
{\bf Behnam Pourhassan\orcidlink{0000-0003-1338-7083}}\footnote{\bf b.pourhassan@du.ac.ir}\\
{\it School of Physics, Damghan University, Damghan 3671645667, Iran}\\
{\it Center for Theoretical Physics, Khazar University, 41 Mehseti Street, Baku, AZ1096, Azerbaijan}\\
{\it Centre for Research Impact \& Outcome, Chitkara University Institute of Engineering and Technology,
Chitkara University, Rajpura, 140401, Punjab, India}\\
{\bf Izzet Sakalli\orcidlink{xxx}}\footnote{\bf izzet.sakalli@gmail.com} \\
{\it Physics Department, Eastern Mediterranean University, Famagusta 99628, North Cyprus via Mersin 10, Turkey}\\
{\bf Jafar Sadeghi\orcidlink{0000-0002-6764-7920}}\footnote{\bf pouriya@ipm.ir}\\
{\it Department of Physics, Faculty of Basic
Sciences, University of Mazandaran\\ P. O. Box 47416-95447, Babolsar, Iran}\\
{\it School of Physics, Damghan University, Damghan 3671645667, Iran}\\
{\it Canadian Quantum Research Center, 204-3002 32 Ave Vernon, BC V1T 2L7, Canada}\\
\end{center}

\vspace{0.3cm}

\begin{abstract}
We investigate the consistency of radiatively corrected inflationary models with both the latest observational data from the Atacama Cosmology Telescope (ACT) combined with Planck 2018 and Baryon Acoustic Oscillation (BAO) measurements, and the theoretical constraints imposed by the swampland program. We systematically test two distinct models against three key swampland conjectures: the further refined de Sitter swampland conjecture (FRDSSC), the scalar weak gravity conjecture (SWGC), and the strong scalar weak gravity conjecture (SSWGC). Model I, based on radiatively corrected Higgs inflation, satisfies the FRDSSC and remains consistent with current observational constraints ($n_s = 0.9743 \pm 0.0034$, $r < 0.038$), but fails to meet the SWGC and SSWGC requirements, indicating limited theoretical compatibility with quantum gravity principles. In contrast, Model II, incorporating radiative corrections with scalar sectors, demonstrates full consistency by satisfying all three swampland conjectures simultaneously while maintaining observational viability. The compatibility of Model II is highly sensitive to the non-minimal coupling $\xi$ and renormalization scales $\mu_b$, with larger values extending the range of swampland-consistent solutions. Our results highlight the critical role of radiative corrections in achieving simultaneous theoretical and observational consistency, and identify Model II as a promising candidate for a fully viable inflationary scenario within the swampland framework. This work provides a methodology for classifying inflationary models based on their swampland compatibility, demonstrating that satisfaction of the FRDSSC alone is insufficient for full theoretical consistency.
\end{abstract}


\tableofcontents

\section{Introduction}\label{isec1}

Recent years have witnessed remarkable advances in our understanding of the early universe, driven by increasingly precise measurements of the Cosmic Microwave Background (CMB)~\cite{1',2'}. Among these, the sixth data release from the ACT, when combined with the \textit{Planck} 2018 data and BAO measurements from the Dark Energy Spectroscopic Instrument (DESI), provides some of the most stringent observational constraints on inflationary cosmology to date~\cite{3',4'}. The joint analysis, referred to as ACT-DR6+Planck18+BAO, yields a scalar spectral index and tensor-to-scalar ratio of $n_s = 0.9743 \pm 0.0034$ and $r < 0.038$~\cite{3',4'}. This result represents an approximate $2\sigma$ upward shift in $n_s$ relative to the \textit{Planck}-only results~\cite{2'}, suggesting a marginally less red-tilted scalar power spectrum. Consequently, inflationary scenarios predicting smaller values of $n_s$ are increasingly disfavored, while models that naturally yield higher $n_s$ values—such as $\alpha$-attractor and related plateau-type models—remain in excellent agreement with the data. These updated bounds have motivated a comprehensive re-evaluation of several classes of inflationary potentials previously considered consistent with \textit{Planck} 2018 constraints~\cite{5',6',7',8',9',10'}.

Beyond inflation itself, these refined constraints significantly impact the post-inflationary epoch, particularly the reheating phase. The effective reheating equation-of-state parameter, $\omega_{\text{eff}}$, now faces much tighter observational limits. As demonstrated in~\cite{15'}, $\alpha$-attractor models characterized by hyperbolic tangent potentials are consistent with the ACT-DR6+Planck18+BAO dataset only if $\omega_{\text{eff}} \geq 0.44$, a lower bound absent in the \textit{Planck}-only analyses. This restriction has far-reaching consequences for early-universe dynamics, as $\omega_{\text{eff}}$ directly influences the amplitude and spectral tilt of primordial gravitational waves during horizon re-entry~\cite{16',17'}. Moreover, reheating parameters affect the generation and evolution of primordial magnetic fields through Faraday induction, making $\omega_{\text{eff}}$ a critical quantity for understanding magnetogenesis~\cite{21',22'}. Thus, ACT's latest results not only refine our picture of inflation but also reshape expectations regarding reheating physics, stochastic gravitational-wave backgrounds, and early cosmic magnetism. These topics are currently being investigated within a variety of scalar-tensor inflationary frameworks~\cite{23'}.

Many recent studies have examined a wide range of inflationary models under the latest ACT observational bounds; a comprehensive list can be found in~\cite{4100,4101,4102,4103,4104,4105,4106,4107,4108,4109,4110,4111,4112,4113,4114}. These investigations have established that precision cosmology is now capable of discriminating between competing theoretical frameworks, making it essential to identify models that satisfy both empirical constraints and fundamental theoretical principles.

In parallel with this rapid observational progress, theoretical advances in high-energy physics have given rise to the \textit{swampland program}, a comprehensive framework aimed at distinguishing low-energy effective field theories that can consistently emerge from an ultraviolet (UV)-complete theory of quantum gravity~\cite{a,b,f,h,i,k,l,o,p,s,v,x,z,bb,cc,dd,ee,ff,hh}. This approach not only tests the internal consistency of such theories but also provides an indirect probe of string theory, which remains our most promising candidate for a fundamental unification of gravity with quantum field theory. The swampland program has emerged as a modern theoretical framework designed to assess the internal coherence of effective field theories that couple to gravity. It serves as a diagnostic tool for probing the logical structure of quantum gravity and, by extension, offers indirect support for string theory. Since direct experimental verification of string theory remains highly challenging due to its characteristic energy scale and mathematical complexity, the swampland framework provides a valuable means of connecting string-inspired principles with cosmological observations.

The program advances along two complementary directions: the \textit{top-down} perspective formulates conjectures rooted in string-theoretic and quantum-gravitational principles, while the \textit{bottom-up} approach evaluates these conjectures by confronting them with explicit cosmological models and observations. Together, these perspectives form a powerful methodology—simultaneously testing the conjectures themselves and constraining viable models of the early universe. This dual approach not only helps evaluate the internal soundness of the swampland conjectures but also serves to identify effective theories that might genuinely arise from a consistent UV-complete framework.

Over time, several conjectures within this framework have been refined to accommodate both theoretical insights and empirical challenges. A prominent example concerns the tension between the original de Sitter (dS) swampland conjecture and conventional single-field slow-roll inflation, which motivated more refined formulations such as the refined de Sitter (RdS) conjecture and the FRDSSC~\cite{2a,3a}. These refinements introduce additional free parameters, thereby expanding the class of inflationary models capable of satisfying swampland constraints. The inclusion of such parameters has proven essential for reconciling theoretical expectations with high-precision cosmological observations, including the latest ACT measurements. The FRDSSC, in particular, unifies earlier constraints into a single extended relation characterized by tunable coefficients and a deformation parameter, offering a wider consistency range for constructing realistic inflationary potentials.

Beyond the FRDSSC, the scalar weak gravity conjecture (SWGC) and SSWGC impose additional theoretical constraints~\cite{6a,7a,8a}. These conjectures generalize the weak gravity conjecture (WGC) from gauge theories to scalar field theories, requiring that scalar-mediated forces dominate over gravitational interactions. Mathematically, this translates into inequalities constraining the hierarchy of derivatives of the scalar potential. While the SWGC provides a baseline consistency condition, the SSWGC imposes more stringent requirements that must hold across the entire field range. Together with the FRDSSC, these conjectures form a comprehensive theoretical toolkit for evaluating the viability of inflationary models from a quantum gravity perspective.

Within this evolving theoretical landscape, models incorporating radiative corrections have attracted considerable attention. Quantum loop effects from Standard Model (SM) particles or beyond-SM fields can significantly modify tree-level inflationary potentials, altering both the shape of the potential and its higher-order derivatives. These radiative corrections play a crucial role in achieving compatibility with both observational constraints and swampland conjectures. In particular, one-loop Coleman-Weinberg corrections introduce logarithmic running that can flatten potentials, suppress tensor-to-scalar ratios, and modify the derivative structure in ways that impact SWGC and SSWGC satisfaction.

Traditional categorizations of inflationary models emphasize potential shapes, underlying symmetries, or dynamical properties. However, a complementary classification based on \textit{swampland compatibility} has become increasingly valuable. The present work adopts this strategy: we systematically examine radiatively corrected inflationary frameworks under multiple swampland conjectures to determine which constructions remain theoretically consistent and empirically supported. In particular, we assess each model's adherence to the FRDSSC, SWGC, and SSWGC. Through this analysis, we aim to identify models that not only satisfy quantum-gravity criteria but also align with the latest CMB constraints from ACT and related datasets~\cite{ii,jj,ll,mm,nn,rr,tt,uu,ww,aaa,bbb,ccc,ddd,eee,hhh,jjj,kkk,lll,mmm,ooo,ppp,qqq,Sadeghi:2023cxh}.

Our investigation focuses on two distinct radiatively corrected inflationary models. Model I examines the radiatively corrected Higgs inflation scenario, where the SM Higgs field serves as the inflaton and quantum corrections from gauge bosons and fermions modify the effective potential. This framework has been extensively studied due to its attractive feature of connecting particle physics with cosmology. However, its consistency with the full suite of swampland conjectures remains an open question. Model II considers a more general setup incorporating a real scalar inflaton non-minimally coupled to gravity, augmented by additional scalar and fermionic fields that introduce competing bosonic and fermionic loop contributions. This richer field content provides additional flexibility for satisfying swampland constraints while maintaining observational viability.

The central aim of this study is to examine how effectively these radiatively corrected inflation frameworks satisfy key swampland conjectures while remaining consistent with the most recent observational limits established by ACT data. By combining theoretical consistency tests with empirical constraints, we seek to determine whether these inflationary scenarios can be considered both observationally viable and theoretically compatible with the principles of quantum gravity. Specifically, we investigate whether radiative corrections can simultaneously improve agreement with ACT observations and enable satisfaction of the stringent SWGC and SSWGC requirements that constrain higher-order derivatives of the potential.

Our analysis reveals a stark distinction between the two models. Model I satisfies the FRDSSC and aligns with current observational constraints from ACT and Planck across various parameter choices. However, when evaluated against the SWGC and SSWGC, the model fails to provide consistent solutions across the relevant parameter space. This indicates that while Model I is observationally viable under the FRDSSC, it remains theoretically limited and cannot be classified as fully consistent within the broader swampland framework. In contrast, Model II demonstrates remarkable consistency on multiple fronts, not only satisfying the FRDSSC but also fulfilling the SWGC and SSWGC requirements. The model's compatibility is particularly sensitive to the values of the non-minimal coupling $\xi$ and the renormalization scales $\mu_b$ and $\mu_f$, with larger values extending the range of consistent solutions. This sensitivity highlights the critical role of radiative corrections and parameter tuning in achieving both theoretical viability and observational accuracy.

It is worth noting that string theory—the primary setting for the swampland program—provides an elegant and comprehensive framework for unifying gravity with the quantum description of matter~\cite{q1,q2,q3,q4,q5,q6,q7,q8,q9,q10,q11,q12,q13,q14,q15,q16,q17,q18,q19,q20,q21,q22,q23,q24}. Its intricate mathematical structure offers a natural context for exploring the geometry and dynamics of the early universe, making it a uniquely powerful tool for investigating how inflationary and post-inflationary physics arise from first principles. The swampland conjectures emerge naturally from string-theoretic considerations, providing a bridge between high-energy fundamental theory and observable cosmological phenomena.

Among the models surveyed in this work, certain realizations of radiatively corrected inflation emerge as particularly promising in achieving simultaneous compatibility with multiple swampland conditions. This makes them valuable candidates for exploring early-universe dynamics and for advancing our understanding of cosmic evolution. Furthermore, the introduction of tunable parameters in the FRDSSC—absent in earlier conjectures—demonstrates that flexibility within theoretical constraints can significantly improve concordance with both fundamental consistency requirements and empirical observations. In this sense, the refined conjectures provide not only a theoretical safeguard against inconsistent inflationary potentials but also a practical bridge connecting quantum-gravity expectations~\cite{Sadeghi:2023cxh,Al-Badawi:2025pjd,Pourhassan:2020zrw,Dengiz:2025jpi,Al-Badawi:2025yqu,Sucu:2025fwa,Sucu:2025dor,Sucu:2025olo,Pourhassan:2022irk} to measurable cosmological phenomena.

The remainder of this paper is organized as follows. Section~\ref{isec2} presents the theoretical framework, introducing the FRDSSC, SWGC, and SSWGC and establishing their mathematical formulations in terms of inflationary observables. Section~\ref{isec3} analyzes Model I (radiatively corrected Higgs inflation), deriving the one-loop effective potential in the Einstein frame (EF), computing slow-roll parameters and inflationary observables, and systematically testing the model against all three swampland conjectures. Section~\ref{isec4} examines Model II (radiatively corrected inflation with scalar-fermion interactions), following a parallel analytical framework and demonstrating full consistency with the FRDSSC, SWGC, and SSWGC. Section~\ref{isec5} concludes with a comprehensive summary of our principal findings, emphasizing the importance of Model II, the limitations of Model I, and the broader implications for inflationary model building. We also outline promising directions for future research, including extensions to warm inflation, multi-field scenarios, and connections with post-inflationary physics.

\section{Swampland Conjectures}\label{isec2}

In this analysis, we focus on identifying inflationary models that remain simultaneously compatible with current cosmological data and consistent with theoretical constraints from quantum gravity. Within the framework of the swampland program, a variety of conjectures have been proposed to delineate the space of low-energy effective theories that can emerge from a UV-complete gravitational background. The refined formulations of these conjectures often introduce adjustable parameters, thereby broadening the range of inflationary potentials that can be considered viable. This added freedom allows for a closer connection between fundamental theory and precision cosmological observations.

The swampland program has emerged as a powerful diagnostic framework for assessing which effective field theories can consistently arise from string theory and quantum gravity. It distinguishes between theories that belong to the \textit{landscape}—those that can be UV-completed—and those confined to the \textit{swampland}, which cannot be consistently embedded in quantum gravity despite appearing viable at low energies. Among the various conjectures within this program, the dS swampland conjecture and its refinements have attracted considerable attention due to their direct implications for inflationary cosmology. A notable example of this development is the FRDSSC, which generalizes earlier no-go conditions while remaining compatible with slow-roll inflationary scenarios. By introducing tunable parameters, the FRDSSC extends the viable parameter space and enables a systematic classification of inflationary models based on their theoretical consistency.

In addition to the FRDSSC, the SWGC and SSWGC impose constraints on the hierarchy of derivatives of the inflationary potential. These conjectures ensure that scalar-mediated forces dominate over gravitational interactions, providing stringent tests for the internal consistency of scalar field theories coupled to gravity. Together, these three conjectures—FRDSSC, SWGC, and SSWGC—form a comprehensive theoretical toolkit for evaluating inflationary models. Our analysis systematically examines two radiatively corrected inflationary scenarios against all three conjectures, assessing their compatibility with both theoretical principles and the latest ACT observational data.

\subsection{FRDSSC}\label{subsec:frdssc}

We begin with the general action describing a set of scalar fields $\varphi^i$ that are minimally coupled to gravity~\cite{2a,3a}:
\begin{equation}\label{eq1}
S = \int d^4x \, \sqrt{|g|} \left[ \frac{1}{2}M_P^2 R - \frac{1}{2} h_{ij} \partial_\mu \varphi^i \partial^\mu \varphi^j - V(\varphi) \right],
\end{equation}
where $M_P$ denotes the reduced Planck mass, $R$ is the Ricci scalar, and $h_{ij}$ represents the field-space metric. This action provides the starting point for analyzing inflationary dynamics within the context of scalar-tensor theories.

The dS and RdS conjectures impose bounds on the potential $V(\varphi)$ whenever $V > 0$~\cite{4a,5a}:
\begin{equation}\label{eq2}
|\nabla V| \geq \frac{c_1}{M_P} V, \hspace{12pt} \text{or} \hspace{12pt} \min(\nabla_i \nabla_j V) \leq -\frac{c_2}{M_P^2} V,
\end{equation}
where $c_1$ and $c_2$ are dimensionless coefficients of order unity. These conditions require that either the potential must be sufficiently steep (first condition) or sufficiently curved (second condition) to avoid inconsistencies with quantum gravity. However, these conditions were found to be in tension with conventional slow-roll inflation, motivating the development of refined versions.

Expressed in terms of potential slow-roll parameters, these inequalities become
\begin{equation}\label{eq3}
\sqrt{2 \epsilon_V} \geq c_1, \hspace{12pt} \text{or} \hspace{12pt} \eta_V \leq -c_2.
\end{equation}

The FRDSSC, proposed by Andriot and Roupec, unifies these two conditions into a single extended relation~\cite{2a,3a}:
\begin{equation}\label{eq4}
\left( M_P \frac{|\nabla V|}{V} \right)^q - a M_P^2 \frac{\min(\nabla_i \nabla_j V)}{V} \geq b,
\end{equation}
where $a,b>0$ satisfy $a+b=1$, and $q>2$. This refinement is particularly suitable for slow-roll inflation, as it offers a wider consistency range for constructing realistic potentials. The introduction of the deformation parameter $q$ and the coefficients $a$ and $b$ provides additional flexibility, allowing a broader class of inflationary models to satisfy swampland constraints. The FRDSSC represents a significant theoretical advancement, as it reconciles the swampland program with observationally viable inflationary scenarios.

The potential slow-roll parameters are defined as
\begin{equation}\label{eq14}
\epsilon_V = \frac{1}{2} \left( \frac{|\nabla V|}{V} \right)^2 = \frac{1}{2} F_1^2, 
\quad \eta_V = \frac{\min(\nabla_i \nabla_j V)}{V} = F_2,
\end{equation}
which allows Eq.~\eqref{eq4} to be rewritten in a compact form:
\begin{equation}\label{eq15}
F_1^q - a F_2 \geq 1 - a.
\end{equation}

The quantities $F_1$ and $F_2$ can be directly related to the inflationary observables as follows:
\begin{equation}\label{eq16}
F_1 = \sqrt{\frac{r}{8}}, \quad 
F_2 = \frac{n_s + \tfrac{3}{8}r - 1}{2},
\end{equation}
where $n_s$ represents the scalar spectral index and $r$ denotes the tensor-to-scalar ratio. This parameterization establishes a direct connection between fundamental theoretical constraints and observable cosmological quantities, enabling quantitative tests of the FRDSSC against CMB data from ACT, Planck, and related experiments.

\subsection{SWGC and SSWGC}\label{subsec:swgc}

A key element within the swampland framework is the WGC, which postulates that in any consistent quantum gravity theory, gravitational interactions must be weaker than other fundamental forces. Initially formulated for gauge interactions, this idea was later generalized to scalar fields by Palti~\cite{6a,7a}. In its scalar version, the SWGC implies that scalar-mediated forces should dominate over gravitational ones, ensuring that the theory remains consistent with quantum gravity principles. This requirement imposes constraints on the higher-order derivatives of the scalar potential, which must satisfy specific hierarchical relations.

Mathematically, the SWGC leads to the inequality
\begin{equation}\label{eq5}
\left( V^{(3)} \right)^2 \geq \frac{\left( V^{(2)} \right)^2}{M_P^2},
\end{equation}
where $V^{(n)}$ denotes the $n$-th derivative of the scalar potential with respect to the canonically normalized inflaton field. This condition ensures that the third derivative of the potential is sufficiently large compared to the second derivative, preventing pathological behaviors in the scalar sector.

A more restrictive form, the SSWGC, further constrains the potential's curvature~\cite{8a}:
\begin{equation}\label{eq6}
2 \left( V^{(3)} \right)^2 - V^{(2)} V^{(4)} \geq \frac{\left( V^{(2)} \right)^2}{M_P^2}.
\end{equation}
Unlike the SWGC, this stronger condition is required to hold for all possible field values, imposing tighter theoretical limits on permissible inflationary potentials. The SSWGC represents one of the most stringent tests within the swampland program, and models that satisfy this conjecture are considered particularly robust from a quantum gravity perspective.

The SWGC and SSWGC play complementary roles to the FRDSSC. While the FRDSSC primarily constrains the first and second derivatives of the potential (related to $\epsilon_V$ and $\eta_V$), the SWGC and SSWGC impose additional constraints on higher-order derivatives, ensuring that the full derivative structure of the potential remains consistent with quantum gravity. Together, these three conjectures provide a comprehensive framework for evaluating the theoretical viability of inflationary models.

In our subsequent analysis, we systematically test both Model I (radiative-corrected Higgs inflation) and Model II (radiatively corrected inflation with scalar-fermion interactions) against all three conjectures. As we demonstrate, Model I satisfies the FRDSSC but fails to meet the SWGC and SSWGC requirements, indicating limited theoretical compatibility. In contrast, Model II achieves full consistency with all three conjectures across appropriate parameter ranges, establishing it as a more theoretically robust inflationary scenario. The parameter sensitivity of these results—particularly the dependence on the non-minimal coupling $\xi$ and the renormalization scales $\mu_b$ and $\mu_f$—provides valuable insights into the structure of viable inflationary theories within the swampland landscape.
\section{Model I}\label{isec3}

In this section, we investigate the framework of quantum-corrected Higgs inflation, following and extending the methodology developed in Refs.~\cite{m1,m2,m3,m4} (see also the comprehensive review in Ref.~\cite{m5,m6}). The Higgs inflation scenario has attracted considerable attention as a compelling mechanism for connecting particle physics with early-universe cosmology. By identifying the SM Higgs field as the inflaton, this framework provides a unified description that simultaneously accounts for electroweak symmetry breaking and inflationary dynamics. However, achieving consistency with observational constraints requires the introduction of a large non-minimal coupling between the scalar field and the Ricci curvature, typically of order $\xi \sim 10^4$.

Quantum corrections play a crucial role in modifying the tree-level predictions of Higgs inflation. At one-loop order, radiative effects from SM particles—particularly the massive gauge bosons and the top quark—induce logarithmic corrections to the effective potential and kinetic terms. These corrections can significantly alter the inflationary observables, such as the scalar spectral index $n_s$ and the tensor-to-scalar ratio $r$, bringing them into closer alignment with CMB measurements from ACT and Planck. Furthermore, radiative corrections influence the higher-order derivatives of the potential, which are directly constrained by the SWGC and SSWGC within the swampland program.

Our analysis proceeds in several stages. First, we introduce the Jordan-frame action and perform the conformal transformation to the EF, where gravity assumes its standard Einstein-Hilbert form. We then derive the one-loop corrected potential in the EF and compute the resulting slow-roll parameters. Next, we test the model against the FRDSSC, demonstrating that it satisfies this refined swampland condition across a range of parameter choices. Finally, we examine the model's compatibility with the SWGC and SSWGC, revealing a critical limitation: while Model I is fully consistent with the FRDSSC and observational data, it fails to satisfy the more stringent SWGC and SSWGC requirements. This selective consistency highlights the hierarchical nature of swampland constraints and underscores the importance of testing inflationary models against the complete suite of swampland conjectures.

\subsection{Jordan-Frame Action and One-Loop Corrections}\label{subsec:jordan_frame}

We begin by considering a general class of scalar-tensor theories characterized by a non-minimal interaction between the scalar sector and gravity. The corresponding action is given by
\begin{equation}
S = \int d^4x \, \sqrt{-g} 
\left[
U(\phi) R 
- \frac{G(\phi)}{2} \nabla_\mu \Phi^a \nabla^\mu \Phi_a
- V(\phi)
\right],
\label{eq:action_general}
\end{equation}
where $\Phi^a$ $(a = 1, \ldots, N)$ denotes an $O(N)$-symmetric multiplet of real scalar fields, and $\delta_{ab}$ is the flat internal metric used to raise and lower internal indices. Specific realizations of this general theory are obtained by choosing appropriate functional forms of $U(\phi)$, $G(\phi)$, and $V(\phi)$, as well as specifying the number of components $N$. To ensure $O(N)$ invariance, all three functions must depend only on the invariant modulus
\begin{equation}
\phi \equiv \sqrt{\Phi^a \Phi_a}.
\end{equation}

In the Higgs inflation scenario, the SM Higgs doublet is identified as the inflaton field. To bring the inflationary predictions into agreement with observations, the model requires a large non-minimal coupling between the scalar and the Ricci scalar of the form $\xi \phi^2 R$, typically with $\xi \sim 10^4$. This strong coupling ensures that the effective Planck mass during inflation is field-dependent and significantly enhanced, leading to a flattened potential in the EF that supports slow-roll dynamics.

Quantum corrections introduce radiative modifications to the tree-level coefficients of the action in Eq.~\eqref{eq:action_general}. To one-loop order, these functions take the form~\cite{m4}
\begin{align}
U(\phi) &= U_{\text{tree}} + U_{\text{1-loop}}
       = \frac{1}{2} (M_P^2 + \xi \phi^2)
       + \frac{\phi^2}{32 \pi^2} C \ln \left( \frac{\phi^2}{\mu^2} \right), 
       \label{eq:U_function} \\[4pt]
V(\phi) &= V_{\text{tree}} + V_{\text{1-loop}}
       = \frac{\lambda}{4} (\phi^2 - \nu^2)^2
       + \frac{\lambda \phi^4}{128 \pi^2} A \ln \left( \frac{\phi^2}{\mu^2} \right), 
       \label{eq:V_function} \\[4pt]
G(\phi) &= G_{\text{tree}} + G_{\text{1-loop}}
       = 1 + \frac{1}{32 \pi^2} E \ln \left( \frac{\phi^2}{\mu^2} \right).
       \label{eq:G_function}
\end{align}

The interaction terms among the Higgs field and the other SM particles can be summarized schematically as
\begin{equation}
\mathcal{L}^{\text{int}}_{\text{SM}} 
= - \sum_{\chi} \frac{\lambda_\chi}{2} \chi^2 \phi^2
  - \sum_A \frac{g_A^2}{2} A_\mu^2 \phi^2
  - \sum_\Psi y_\Psi \phi \bar{\Psi} \Psi,
\label{eq:SM_int}
\end{equation}
where $\lambda_\chi$, $g_A$, and $y_\Psi$ correspond to scalar, gauge, and Yukawa couplings, respectively. The leading contributions to the effective potential arise from the heaviest SM fields, yielding
\begin{equation}
m_W^2 = \frac{g^2}{4} \phi^2, 
\qquad 
m_Z^2 = \frac{g^2 + g'^2}{4} \phi^2, 
\qquad 
m_t^2 = \frac{y_t^2}{2} \phi^2.
\label{eq:mass_relations}
\end{equation}

At high field values, the logarithmic coefficients $A$, $C$, and $E$ in Eqs.~\eqref{eq:U_function}–\eqref{eq:G_function} receive their dominant contributions from Goldstone boson loops. Neglecting graviton loop effects—which are suppressed by the large effective Planck mass,
\begin{equation}
M_{P, \text{eff}} = \sqrt{M_P^2 + \xi \phi^2},
\end{equation}
and performing an expansion in powers of $1/\xi$, one obtains
\begin{align}
A &= \frac{3}{8\lambda} 
\Big[ 2g^4 + (g^2 + g'^2)^2 - 16y_t^4 \Big]
+ 6\lambda + \mathcal{O}(\xi^{-2}), \nonumber \\
C &= 3 \xi \lambda + \mathcal{O}(\xi^0), \qquad
E = \mathcal{O}(\xi^{-2}).
\label{eq:ACE}
\end{align}

These expressions reveal the structure of radiative corrections and their dependence on SM parameters. The coefficient $A$ encodes contributions from gauge bosons and the top quark, while $C$ is dominated by the non-minimal coupling $\xi$. The suppression of $E$ at order $\xi^{-2}$ ensures that kinetic corrections remain subdominant in the large-$\xi$ regime.

\subsection{Einstein-Frame Transformation and Effective Potential}\label{subsec:einstein_frame}

The Jordan-frame action in Eq.~\eqref{eq:action_general} can be mapped to an EF representation—where gravity is minimally coupled to a scalar field—by performing a conformal transformation of the metric and a field redefinition:
\begin{align}
g_{\mu\nu} &\rightarrow \bar{g}_{\mu\nu} = \frac{2U(\phi)}{M_P^2} g_{\mu\nu}, 
\label{eq:conformal_transform} \\[3pt]
\left( \frac{d\chi}{d\phi} \right)^2 
&= \frac{M_P^2}{2} 
\left( 
\frac{G(\phi) U(\phi) + 3 [U'(\phi)]^2}{U^2(\phi)} 
\right),
\label{eq:field_redef} \\[3pt]
V(\chi) &= 
\left( \frac{M_P^2}{2} \right)^2 
\frac{V(\phi)}{U^2(\phi)} 
\Bigg|_{\phi = \phi(\chi)}.
\label{eq:Einstein_potential}
\end{align}

In the regime of large non-minimal coupling, $\xi \gg 1$, we focus on inflationary dynamics at field values satisfying
\begin{equation}
\phi^2 \gg \frac{M_P^2}{\xi} \gg \nu^2.
\label{eq:large_phi}
\end{equation}

In this limit, we assume the following smallness conditions hold:
\begin{equation}
\frac{M_P^2}{\xi \phi^2} \ll 1, 
\qquad 
\frac{A}{32\pi^2} \ll 1,
\label{eq:smallness}
\end{equation}
and expect that all other combinations of loop coefficients (e.g., $B, C, D, E, F$) satisfy similar relations:
\begin{equation}
\frac{1}{32\pi^2} (B, C, D, E, F) \ll 1.
\end{equation}

These assumptions guarantee the validity of the slow-roll approximation within the EF description. Combining Eqs.~\eqref{eq:U_function}–\eqref{eq:field_redef}, we can express the canonical field $\chi$ in terms of the original Higgs field $\phi$ as
\begin{equation}
\chi \simeq 
\sqrt{\frac{3}{2}} M_P 
\left( 
1 + \frac{C}{16 \pi^2 \xi} 
\right)
\ln \left( \frac{\xi \phi^2}{M_P^2} \right).
\label{eq:chi_phi_relation}
\end{equation}

Inverting Eq.~\eqref{eq:chi_phi_relation} yields
\begin{equation}
\phi \simeq 
\frac{1}{\sqrt{\xi}} 
\exp\left( 
\frac{\chi}{\sqrt{6} M_P}
\right)
- 
\frac{C\exp\left( 
\frac{\chi}{\sqrt{6} M_P}
\right)}{16 \pi^2 \xi^{3/2}}  
\frac{\chi}{\sqrt{6} M_P}.
\label{eq:phi_chi_inverse}
\end{equation}

Finally, substituting Eq.~\eqref{eq:phi_chi_inverse} into the EF potential in Eq.~\eqref{eq:Einstein_potential}, we obtain a formulation suitable for computing cosmological observables such as the spectral index and tensor-to-scalar ratio. This representation makes it possible to investigate quantum-corrected Higgs inflation within a consistent theoretical and observational framework.

\subsection{Inflationary Observables and Slow-Roll Dynamics}\label{subsec:slow_roll}

We now express the inflationary potential in the EF in terms of the canonically normalized field $\chi$. Following Ref.~\cite{m5,m6}, the one-loop corrected potential takes the approximate form
\begin{equation}
V(\chi) \simeq 
\frac{\lambda M_P^4}{4 \xi^2}
\left[ 
1 - 2 e^{-\sqrt{2/3} \, \chi/M_P}
+ \frac{A_I \chi}{16\sqrt{6} \pi^2 M_P}
\right],
\label{eq:V_EF}
\end{equation}
where $A_I = A - 12\lambda$ represents the inflationary anomalous scaling, incorporating the quantum corrections from the Goldstone sector via the coefficient $C$. Equation~\eqref{eq:V_EF} provides the radiatively corrected potential relevant for Higgs inflation in the EF representation.

Using the potential in Eq.~\eqref{eq:V_EF}, the slow-roll parameters $\epsilon$ and $\eta$ are readily obtained as
\begin{equation}
\epsilon \simeq 
\frac{4}{3} e^{-2\sqrt{2/3} \, \chi/M_P}
\left( 
1 + \frac{A_I}{64\pi^2} e^{\sqrt{2/3} \, \chi/M_P}
\right)^2
\simeq 
\frac{4 M_P^4}{3 \xi^2 \phi^4}
\left( 
1 + \frac{\phi^2}{\phi_I^2}
\right)^2,
\label{eq:epsilon}
\end{equation}
\begin{equation}
\eta \simeq 
-\frac{4}{3} e^{-2\sqrt{2/3} \, \chi/M_P}
\simeq 
-\frac{4 M_P^4}{3 \xi^2 \phi^4},
\label{eq:eta}
\end{equation}
where we have defined
\begin{equation}
\phi_I^2 \equiv \frac{64\pi^2}{A_I} \xi.
\end{equation}

Following the treatment in Ref.~\cite{m5,m6}, the number of e-folds during inflation, beginning at $\phi$ and ending at $\phi_{\text{end}}$, is given by
\begin{equation}
N = -\int_{\phi}^{\phi_{\text{end}}} d\phi' \, \frac{3H^2(\phi')}{F(\phi')},
\label{eq:e_folds}
\end{equation}
where the function $F(\phi)$ is defined as
\begin{equation}
F(\phi) = 2VU' - V'U.
\end{equation}

Under the slow-roll approximation and the conditions of Eq.~\eqref{eq:smallness}, Eq.~\eqref{eq:e_folds} yields
\begin{equation}
N \simeq 
\frac{48\pi^2}{A_I} \ln \left( 1 + \frac{\phi^2}{\phi_I^2} \right),
\label{eq:N_approx}
\end{equation}
whose inverse relation can be written as
\begin{equation}
\frac{\phi^2}{\phi_I^2} \simeq e^x - 1, 
\qquad x \equiv \frac{A_I N}{48\pi^2}.
\label{eq:phi_N}
\end{equation}

The dimensionless power spectrum of scalar perturbations at horizon crossing $(k = aH)$ is given by
\begin{equation}
\Delta^2(k) = \frac{V(\chi)}{24\pi^2 M_P^4 \epsilon(\chi)}.
\label{eq:power_spectrum}
\end{equation}

Substituting Eqs.~\eqref{eq:V_EF} and~\eqref{eq:epsilon}, we obtain
\begin{equation}
\Delta^2(k) = \frac{N^2 \lambda}{72\pi^2 \xi^2}
\left( \frac{e^x - 1}{x e^x} \right)^2.
\label{eq:Delta2}
\end{equation}

Employing the observed normalization $\Delta^2(k) \simeq 2.1 \times 10^{-9}$ and setting $N = 60$, we infer the approximate relation
\begin{equation}
\frac{\lambda}{\xi^2} \simeq 4.145 \times 10^{-10}
\left( \frac{x e^x}{e^x - 1} \right)^2.
\label{eq:lambda_xi}
\end{equation}

The scalar spectral index and tensor-to-scalar ratio are expressed in the standard slow-roll form as
\begin{equation}
n_s = 1 - 6\epsilon(\chi) + 2\eta(\chi), 
\qquad r = 16\epsilon(\chi).
\label{eq:ns_r}
\end{equation}

From Eqs.~\eqref{eq:epsilon}–\eqref{eq:phi_N}, these quantities become
\begin{equation}
n_s = 1 - \frac{2}{N} \frac{x}{e^x - 1},
\qquad
r = \frac{12}{N^2} \left( \frac{x e^x}{e^x - 1} \right)^2.
\label{eq:ns_r_explicit}
\end{equation}

In the small-$x$ limit $(x \ll 1)$, one finds
\begin{equation}
n_s \simeq 1 - \frac{2}{N}, 
\qquad 
r \simeq \frac{12}{N^2},
\label{eq:ns_r_small_x}
\end{equation}
which reproduce the predictions of both the quadratic potential $V(\phi) = m^2 \phi^2/2$ and the Starobinsky-type $f(R) = R + R^2/(6M^2)$ inflationary model. Hence, the radiatively corrected Higgs inflation scenario exhibits a tensor-to-scalar ratio substantially smaller than those predicted by chaotic inflation models, while preserving the same leading-order spectral behavior.

\subsection{Theoretical Consistency: FRDSSC Analysis}\label{subsec:frdssc_analysis}

To evaluate the theoretical viability of the model, we test it against the FRDSSC by employing the potential derived in the EF $(V(\chi))$. Substituting the potential into Eqs.~\eqref{eq15}–\eqref{eq16}, we compute the quantities $F_1$ and $F_2$, which encode the first and second derivatives of the potential, respectively. The results indicate that while the model does not satisfy the original dS conjecture, it can comply with the FRDSSC for specific parameter choices of $a$, $b$, and $q$.

For $N = 60$, $q = 3$, $r = 0.0054$, and $n_s = 0.9743$, the relevant inequality takes the schematic form
\begin{equation}\label{eq:F1F2_theory}
F_1 = 0.02598076211, \qquad F_2 = -0.01183750000.
\end{equation}
The standard dS bound is violated. For the FRDSSC, we have
\begin{equation}\label{eq:frdssc_theory}
a\geq 0.9882836552,
\end{equation}
which can be satisfied for appropriate $a$, $b$, and deformation parameter $q$. Hence, the consistent theoretical configuration is found to be
\begin{equation}\label{eq:ab_theory}
a = 0.997, \qquad b = 1 - a = 0.003,
\end{equation}
confirming the compatibility of the model with the FRDSSC framework.
\subsection{Observational Constraints: ACT and Planck Data}\label{subsec:obs_constraints}

We now compare the theoretical predictions with observational data. Using the latest results from the ACT, we take $n_s = 0.9743$ and $r < 0.038$ as representative values. Substituting these into Eq.~\eqref{eq16}, we obtain
\begin{equation}\label{eq:F1F2_ACT}
F_1 = 0.06123724358, \qquad F_2 = -0.0072250000.
\end{equation}

As before, the standard dS condition is not fulfilled. For the FRDSSC, we can calculate
\begin{equation}\label{eq:frdssc_ACT}
0 \leq -0.9997703603 + 1.007225000 \, a.
\end{equation}

Nevertheless, by adopting the FRDSSC with parameter $q = 3$, Eq.~\eqref{eq15} yields a viable configuration satisfying
\begin{equation}\label{eq:a_ACT}
0.9925988337 \leq a,
\end{equation}
leading to
\begin{equation}\label{eq:ab_ACT}
a = 0.998, \qquad b = 1 - a = 0.002.
\end{equation}

This confirms the consistency of the model with empirical observations within the FRDSSC framework. To quantify the difference between theoretical and observationally constrained parameters, we define
\begin{equation}\label{eq:Delta_ACT}
\Delta a = 0.001, \qquad \Delta b = 0.001.
\end{equation}

For higher-$q$ formulations of the FRDSSC, significantly narrow the allowed parameter space, leading to near alignment between theoretical and observational constraints.

We also confront the model predictions with Planck observational data. Using the scalar spectral index $n_s =0.9649\pm0.0042$ and tensor-to-scalar ratio $r < 0.064$, we obtain the observational counterparts of the theoretical derivatives:
\begin{equation}\label{eq:F1F2_Planck}
F_1 = 0.07905694147, \qquad F_2 = -0.0081750000.
\end{equation}

As before, the strict dSSC condition is violated. Nevertheless, the FRDSSC condition can still be satisfied. For $q = 3$, it requires
\begin{equation}\label{eq:a_Planck}
0.9914011894 \leq a.
\end{equation}

A consistent parameter set satisfying the FRDSSC is
\begin{equation}\label{eq:ab_Planck}
a = 0.998, \qquad b = 1 - a = 0.002,
\end{equation}
demonstrating that the model remains consistent with current observational limits. To quantify deviations between theoretical and observationally determined configurations, we define
\begin{equation}\label{eq:Delta_Planck}
\Delta a = 0.001, \qquad \Delta b = 0.001.
\end{equation}

We can perform similar calculations for these conjectures with respect to other free parameters such as $N = 56$, $q = 6.5$, and $r = 0.05$. In this case, the dSSC condition is violated, while the FRDSSC introduces a softened, combined condition that restores consistency, requiring $a \geq 0.9883009193$. For $q = 6.5$, a consistent parameter set is $a = 0.997$ and $b = 1 - a = 0.003$, ensuring compliance with the FRDSSC and theoretical consistency of the inflationary setup. The same procedure can be applied to other observational datasets, such as ACT and Planck. To quantify the deviations between theoretical predictions and observationally determined configurations, we define $\Delta a = 0.001$ and $\Delta b = 0.001$.

Similarly, for $N = 56$, $q = 2.01$, and $r = 0.05$, the dSSC condition is violated and the FRDSSC condition that ensures consistency requires $a \geq 0.9876577967$. For $q = 2.01$, a consistent parameter set is $a = 0.997$ and $b = 1 - a = 0.003$, again satisfying the FRDSSC. The deviations between theoretical and observationally determined configurations are $\Delta a = 0.001$ and $\Delta b = 0.001$.

The information regarding the theoretical and observational viability of the model, as well as its compatibility with the various swampland conjectures, is summarized in Table~\ref{tab:frdssc_constraints}. This table demonstrates that Model I satisfies the FRDSSC across all considered scenarios—theoretical tests, ACT observations, and Planck data—with only minor variations in the required values of $a$ and $b$. The consistency of these parameter sets across different observational datasets confirms the robustness of the FRDSSC as a viable swampland criterion for radiatively corrected Higgs inflation.

\begin{table}[h!]
\centering
\caption{Theoretical and Observational Constraints on Model I Parameters with FRDSSC Compliance (Radiative-Corrected Higgs Inflation)}
\label{tab:frdssc_constraints}
\resizebox{\textwidth}{!}{%
\begin{tabular}{ccccccccccc}
\toprule
\textbf{Scenario / Dataset} & \textbf{N} & \textbf{q} & \textbf{r} & \textbf{$n_s$} & \textbf{$F_1$} & \textbf{$F_2$} & \textbf{dSSC Status} & \textbf{FRDSSC Condition} & \textbf{(a, b)} & $\Delta a, \Delta b$ \\
\midrule
Theoretical Test       & 60    & 3     & 0.0054 & 0.9743   & 0.02598076211 & -0.01183750000 & Violated & $a \ge 0.9882836552$ & (0.997, 0.003) & 0.001, 0.001 \\
Observational ACT      & 60    & 3     & $<0.038$ & 0.9743 & 0.06123724358 & -0.0072250000  & Violated & $a \ge 0.9925988337$ & (0.998, 0.002) & 0.001, 0.001 \\
Observational Planck   & 60    & 3     & $<0.064$  & 0.9649 & 0.07905694147 & -0.0081750000 & Violated & $a \ge 0.9914011894$ & (0.998, 0.002) & 0.001, 0.001 \\
Theoretical II & 56 & 6.5   & 0.0054   & 0.9743   &    0.02598076211    & -0.01183750000   & Violated & $a \ge 0.9883009193 $ & (0.997, 0.003) & 0.001, 0.001 \\
Theoretical III & 56 & 2.01  & 0.0054  & 0.9743   &     0.02598076211   & -0.01183750000   & Violated & $a \ge 0.9876577967$ & (0.997, 0.003) & 0.001, 0.001 \\
\bottomrule
\end{tabular}%
}
\end{table}

\bigskip
\noindent
\textbf{Notes:} 
\begin{itemize}
    \item dSSC is violated in all cases.
    \item FRDSSC conditions are satisfied for specific choices of $a$, $b$, $q$.
    \item $\Delta a$, $\Delta b$ quantify the deviation between theoretical and observationally constrained parameters.
\end{itemize}

\subsection{SWGC and SSWGC for Model I}\label{subsec:swgc_model1}

Inflationary models driven by scalar fields serve as excellent testing grounds for the swampland conjectures, as they must simultaneously reproduce the observed values of the scalar spectral index $n_s$ and tensor-to-scalar ratio $r$, while also satisfying consistency with quantum gravity. Radiative corrections and non-minimal couplings can strongly influence higher-order derivatives of the potential, making the SWGC and SSWGC particularly sensitive probes. In this context, the swampland framework distinguishes between effective field theories that can emerge from consistent UV completions and those that belong to the so-called swampland.

Our focus is on testing the radiatively corrected inflationary model against the consistency conditions in Eqs.~\eqref{eq5} and~\eqref{eq6}, using the EF potential $(V(\chi))$ and the parameter ranges established above. By substituting the explicit form of the EF potential into these inequalities, we can determine whether Model I satisfies the SWGC and SSWGC requirements.

\subsubsection{Formulation of the Constraints}\label{subsubsec:constraints_formulation}

By substituting the explicit form of the EF potential into Eqs.~\eqref{eq5} and~\eqref{eq6}, the SWGC and SSWGC impose the conditions:
\begin{equation}\label{eq:SWGC_Model1}
0 \leq 
- \frac{M_{P}^{2} \, \exp\Big(-\frac{2 \sqrt{6} \, \chi}{3 M_{P}}\Big) \, \lambda^{2} \, (-2 + 3 M_{P})}{27 \, \xi^{4}}
\end{equation}
and
\begin{equation}\label{eq:SSWGC_Model1}
0 \leq 
- \frac{M_{P}^{2} \, \omega^{2} \, e^{-\frac{2 \sqrt{6} \, \chi}{3 M_{P}}} \left(M_{P}^{2} - \frac{2}{3}\right)}{9}.
\end{equation}

The fulfillment of these relations ensures that the inflationary potential remains compatible with both theoretical consistency and the refined swampland bounds. However, as we demonstrate below, Model I fails to satisfy these conditions within the physically relevant parameter space.

\begin{figure}[h!]
 \begin{center}
 \subfigure[SWGC for Model I]{
 \includegraphics[height=6cm,width=8cm]{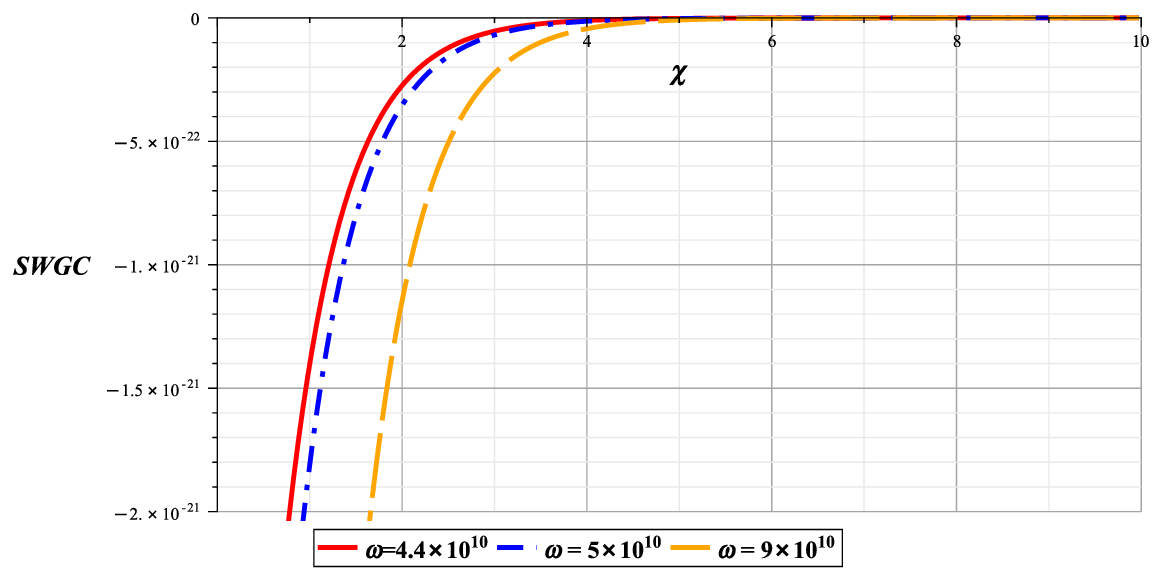}
 \label{fig:swgc_model1}}
 \subfigure[SSWGC for Model I]{
 \includegraphics[height=6cm,width=8cm]{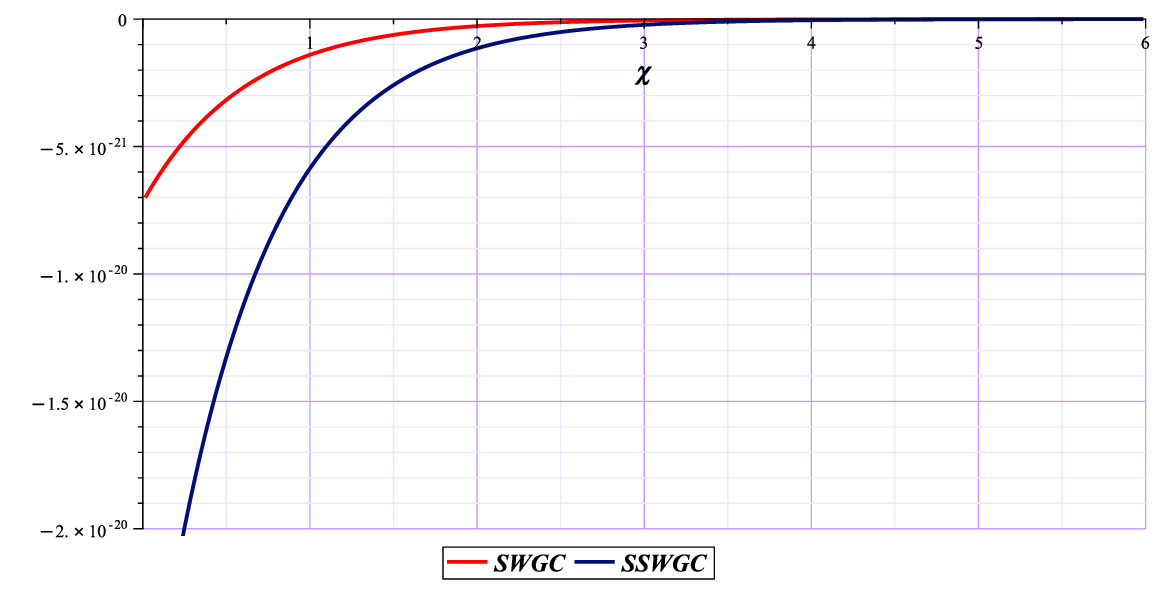}
 \label{fig:sswgc_model1}}
  \caption{SWGC and SSWGC for Model I with respect to $\omega = 9 \times 10^{-10}$, $M_P = 1$. The plots demonstrate that no compatible intervals exist that simultaneously satisfy either the SWGC or the SSWGC for the radiatively corrected Higgs inflation model.}
 \label{fig:model1_wgc}
 \end{center}
 \end{figure}

\subsubsection{SWGC and SSWGC Analysis for Model I}\label{subsubsec:swgc_analysis_model1}

Having established that the radiatively corrected inflationary model (Model I) fully satisfies the FRDSSC and aligns with the latest observational data, we now extend our analysis to the remaining conjectures in the swampland program, namely SWGC and SSWGC. These conjectures impose additional theoretical constraints on scalar field potentials and are crucial for assessing the overall consistency of any inflationary scenario within the swampland framework.

By examining Eqs.~\eqref{eq:SWGC_Model1} and~\eqref{eq:SSWGC_Model1} and considering the model parameters $\omega = 9 \times 10^{-10}$ and $M_P = 1$, we perform a detailed analysis of the parameter space for Model I. Within the intervals explored, our computations reveal that \textbf{no compatible intervals exist} that simultaneously satisfy either the SWGC or the SSWGC. This indicates that, although the model is fully consistent with the FRDSSC, it fails to meet the additional theoretical requirements imposed by the SWGC and SSWGC.

Figure~\ref{fig:model1_wgc} provides a visual confirmation of this outcome: while the FRDSSC-allowed region remains fully accessible within the considered parameter space, the regions corresponding to SWGC and SSWGC compatibility do not intersect with the model's viable configurations. This clearly demonstrates that, for the radiatively corrected inflationary potential, the FRDSSC emerges as the only swampland conjecture that can be satisfied in its entirety.

Consequently, although Model I shows excellent agreement with observational constraints under the FRDSSC, its inability to satisfy the SWGC and SSWGC highlights a limitation from the broader swampland perspective. Specifically, while FRDSSC allows for theoretically and observationally consistent inflationary configurations, the model cannot serve as a universal framework for inflationary scenarios that are fully compliant with \textit{all} swampland conjectures.

In summary, the analysis underscores a key feature of the radiatively corrected Higgs inflation model: it is robust under the FRDSSC, ensuring theoretical viability and observational consistency, but it does not fulfill the full suite of swampland criteria. This selective consistency suggests that while the FRDSSC provides a meaningful and attainable benchmark for inflationary model building, additional conjectures such as the SWGC and SSWGC impose stricter constraints that are not automatically satisfied, thereby narrowing the class of fully viable models within the swampland paradigm. The failure of Model I to satisfy the SWGC and SSWGC motivates the investigation of alternative inflationary frameworks, such as Model II, which we analyze in the following section.

\section{Model II}\label{isec4}

In this section, we investigate an alternative inflationary framework that incorporates radiative corrections arising from both scalar and fermionic sectors. Unlike Model I, which focuses on the radiatively corrected Higgs inflation scenario, Model II considers a real scalar inflaton $\phi$ non-minimally coupled to gravity, augmented by additional scalar fields and right-handed neutrinos. This framework naturally emerges in unified inflaton-dark matter scenarios and provides a pathway for connecting inflation with post-inflationary physics, particularly reheating and leptogenesis. The model's richer field content introduces more complex radiative corrections, which significantly alter the structure of the effective potential and its higher-order derivatives.

A key feature of Model II is the interplay between bosonic and fermionic loop contributions. The bosonic corrections, primarily from the portal coupling between the inflaton and an additional scalar field $H$, tend to enhance the curvature of the potential and increase the tensor-to-scalar ratio $r$. Conversely, fermionic corrections from right-handed neutrinos flatten the potential and suppress $r$. This competition between opposing radiative effects provides a mechanism for fine-tuning the inflationary predictions to match observational constraints from ACT and Planck.

Our analysis demonstrates that Model II achieves full compatibility with all three swampland conjectures: FRDSSC, SWGC, and SSWGC. This stands in stark contrast to Model I, which satisfies only the FRDSSC while failing the more stringent SWGC and SSWGC tests. The successful satisfaction of all swampland criteria in Model II is highly parameter-dependent, with the non-minimal coupling $\xi$ and the renormalization scales $\mu_b$ and $\mu_f$ playing crucial roles. Higher values of $\xi$ and $\mu_b$ extend the parameter space over which the model remains consistent with swampland bounds, highlighting the importance of careful parameter selection in constructing theoretically viable inflationary models.

We organize our discussion as follows. First, we introduce the Jordan-frame action and derive the one-loop corrected effective potential. Next, we perform the conformal transformation to the EF and obtain the radiatively corrected potential in terms of the canonical field $\chi$. We then compute the slow-roll parameters and test the model against the FRDSSC, demonstrating full consistency across various parameter choices and observational datasets. Finally, we examine the model's compatibility with the SWGC and SSWGC, presenting detailed numerical results and visualizations that confirm Model II as a fully viable inflationary scenario within the swampland program.

\subsection{Jordan-Frame Action with Scalar and Fermionic Fields}\label{subsec:jordan_action_model2}

We begin the study in the Jordan frame, where the tree-level action consists of a real scalar inflaton $\phi$ non-minimally coupled to gravity, augmented by additional scalar and fermionic fields~\cite{n1,n2,n3}. The complete action is written as
\begin{align}
S_J = \int d^4x \sqrt{-g} \Bigg[
& -\frac{1}{2}\left(M_P^2 + \xi_\phi \phi^2 + \xi_H H^\dagger H \right) R
\label{eq:Jordan_action} \\[4pt]
& + \frac{1}{2}(\partial_\mu \phi)(\partial^\mu \phi)
 - \frac{1}{2} m_\phi^2 \phi^2
 - \frac{\lambda_\phi}{4!}\phi^4
\label{eq:Jordan_phi} \\[4pt]
& + (\partial_\mu H)^\dagger(\partial^\mu H)
 - (\lambda_{\phi H}\phi^2 - m_H^2) H^\dagger H
 - \lambda_H (H^\dagger H)^2
\label{eq:Jordan_H} \\[4pt]
& + i \bar{N}\gamma^\mu \partial_\mu N
 - \frac{1}{2}\big(y_\phi \phi + m_N\big)\bar{N}^c N + \cdots
\Bigg].
\label{eq:Jordan_N}
\end{align}

Here $M_P$ is the reduced Planck mass, and $\xi_\phi$ ($\xi_H$) denotes the non-minimal coupling between $\phi$ ($H$) and the Ricci scalar $R$. The quartic couplings $\lambda_\phi$ and $\lambda_H$ are positive, while $\lambda_{\phi H}$ quantifies the Higgs-inflaton portal interaction, a feature common to unified inflaton-dark sector frameworks~\cite{n4} and other beyond-SM scenarios~\cite{n5,n6}. 

The right-handed neutrino field $N$ is an SM singlet and couples to the inflaton through the Yukawa term $y_\phi \phi \bar{N}^c N$. The Majorana mass combination $(y_\phi \phi + m_N)\bar{N}^c N$ supports a Type-I seesaw mechanism, simultaneously enabling reheating through lepton-number-violating processes, thereby providing a link to baryogenesis via leptogenesis. During inflation, the inflaton field $\phi$ acquires large values, inducing heavy effective masses for the scalar $H$ through the portal term $\lambda_{\phi H}\phi^2 H^\dagger H$. Consequently, $H$ is dynamically stabilized at the origin, reducing the dynamics effectively to a single-field system governed by $\phi$. Nonetheless, the presence of $H$ and $N$ significantly impacts the inflaton potential through radiative corrections of Coleman-Weinberg type.

\subsection{One-Loop Corrected Potential in the Jordan Frame}\label{subsec:one_loop_jordan}

Quantum fluctuations of the scalar and fermionic sectors induce loop corrections that modify the classical potential and alter inflationary predictions. The one-loop effective potential in the Jordan frame can be expressed as
\begin{align}
V_J(\phi) =
\frac{1}{2} m_\phi^2 \phi^2 
+ \frac{\lambda_\phi}{4!}\phi^4
+ \frac{1}{64\pi^2} \Bigg[
& \left(m_\phi^2 + \frac{\lambda_\phi}{2}\phi^2\right)^2
\ln \!\left(\frac{m_\phi^2 + \frac{\lambda_\phi}{2}\phi^2}{\mu^2}\right)
\label{eq:CW_boson} \\[4pt]
& + 2N_H (\lambda_{\phi H}\phi^2 - m_H^2)^2
\ln \!\left(\frac{\lambda_{\phi H}\phi^2 - m_H^2}{\mu^2}\right)
\label{eq:CW_Higgs} \\[4pt]
& - 2N_N (y_\phi \phi + m_N)^4
\ln \!\left(\frac{(y_\phi \phi + m_N)^2}{\mu^2}\right)
\Bigg],
\label{eq:CW_fermion}
\end{align}
where $N_H$ is the number of scalar degrees of freedom in $H$, and $N_N=3$ corresponds to the number of right-handed neutrinos.

For the inflationary regime $\lambda_\phi \phi^2 \gg m_\phi^2$, $\lambda_{\phi H}\phi^2 \gg m_H^2$, and $y_\phi \phi \gg m_N$, the potential simplifies to the asymptotic form
\begin{equation}
V_J(\phi) \simeq
\frac{\lambda_\phi}{4!}\phi^4
+ \kappa_b \phi^4 \ln\!\left(\frac{\phi}{\mu_b}\right)
- \kappa_f \phi^4 \ln\!\left(\frac{\phi}{\mu_f}\right),
\label{eq:VJ_approx}
\end{equation}
with loop coefficients
\begin{equation}
\kappa_b = \frac{N_H \lambda_{\phi H}^2}{16\pi^2},
\qquad
\kappa_f = \frac{N_N y_\phi^4}{16\pi^2},
\label{eq:loop_coeffs}
\end{equation}
and characteristic renormalization scales $\mu_b = \mu/\sqrt{\lambda_{\phi H}}$ and $\mu_f = \mu / y_\phi$.

The coefficient $\kappa_b$ encodes the bosonic loop contributions from the portal scalar $H$, while $\kappa_f$ represents fermionic corrections from right-handed neutrinos. The logarithmic structure of Eq.~\eqref{eq:VJ_approx} reflects the running of the effective coupling constants with field strength, a characteristic feature of radiatively corrected inflationary potentials. The opposing signs of $\kappa_b$ and $\kappa_f$ terms indicate that bosonic and fermionic contributions compete, providing a natural mechanism for balancing the potential's shape and achieving observationally viable predictions.

\subsection{Transition to the Einstein Frame}\label{subsec:einstein_transformation}

To analyze inflation in a frame with canonical gravity, we apply the conformal rescaling~\cite{n7}
\begin{equation}
g_{\mu\nu} \rightarrow g^{E}_{\mu\nu} = \Omega^2(\phi) \, g_{\mu\nu},
\qquad
\Omega^2(\phi) = 1 + \frac{\xi_\phi \phi^2}{M_P^2},
\label{eq:conformal_transform_model2}
\end{equation}
which removes the non-minimal coupling and converts the gravitational part into the standard Einstein-Hilbert form. The EF action becomes
\begin{equation}
S_E = \int d^4x \sqrt{-g_E}
\left[
-\frac{1}{2} M_P^2 R_E + \mathcal{L}_E(\phi,H,N)
\right].
\label{eq:Einstein_action_model2}
\end{equation}

However, this rescaling introduces a non-canonical kinetic term for $\phi$. To restore canonical normalization, we define a new field $\chi$ via
\begin{equation}
\left(\frac{d\chi}{d\phi}\right)^2 =
\frac{1 + \frac{\xi_\phi \phi^2}{M_P^2} + \frac{6\xi^2_\phi\phi^2}{M_P^2}}{\left(1 + \frac{\xi_\phi \phi^2}{M_P^2}\right)^2}.
\label{eq:field_redef_model2}
\end{equation}

In the large-field limit $\phi \gg M_P/\sqrt{\xi_\phi}$, this relation integrates to
\begin{equation}
\chi(\phi) \simeq 
\sqrt{\frac{3}{2}}\, M_P
\ln\!\left(\frac{\xi_\phi \phi^2}{M_P^2}\right),
\qquad
\phi(\chi) \simeq 
\frac{M_P}{\sqrt{\xi_\phi}}
\exp\!\left(\frac{\chi}{\sqrt{6}M_P}\right).
\label{eq:field_relation_model2}
\end{equation}

This field redefinition is crucial for obtaining a canonical kinetic term in the EF, which is necessary for computing the slow-roll parameters and inflationary observables using standard techniques. The exponential relationship between $\chi$ and $\phi$ in Eq.~\eqref{eq:field_relation_model2} implies that the potential in the EF will exhibit plateau-like behavior, a characteristic feature that facilitates slow-roll inflation and leads to predictions consistent with CMB observations.

\subsection{Radiatively Corrected Einstein-Frame Potential}\label{subsec:radiative_potential_model2}

The potential in the EF is obtained by rescaling the Jordan-frame potential by the conformal factor:
\begin{equation}
V_E(\phi) = \frac{V_J(\phi)}{\Omega^4(\phi)}.
\label{eq:VE_general_model2}
\end{equation}

Using Eq.~\eqref{eq:VJ_approx}, the one-loop corrected EF potential becomes~\cite{n7}
\begin{equation}
V_E(\phi) =
\frac{1}{\big(1 + \xi_\phi \phi^2 / M_P^2\big)^2}
\left[
\frac{\lambda_\phi}{4!}\phi^4
+ \kappa_b \phi^4 \ln\!\left(\frac{\phi}{\mu_b}\right)
- \kappa_f \phi^4 \ln\!\left(\frac{\phi}{\mu_f}\right)
\right].
\label{eq:VE_approx_model2}
\end{equation}

The bosonic loop contribution $\kappa_b$ enhances the curvature of the potential and typically leads to larger values of the tensor-to-scalar ratio, while fermionic corrections $\kappa_f$ act oppositely, flattening the potential and suppressing $r$. The interplay between these two effects crucially determines the inflationary predictions, as we will quantify below.

Inflation is governed by the slow-roll parameters, defined in terms of the EF potential and the field redefinition function:
\begin{align}
\epsilon(\phi) &=
\frac{M_P^2}{2}
\left(
    \frac{V_E'}{V_E\, \chi'}
\right)^2,
\label{eq:epsilon_def_model2} \\[6pt]
\eta(\phi) &=
M_P^2 \left[
    \frac{V_E''}{V_E\, (\chi')^2}
    - \frac{V_E' \chi''}{V_E\, (\chi')^3}
\right],
\label{eq:eta_def_model2} \\[6pt]
\xi^2(\phi) &=
M_P^4
\left(
    \frac{V_E'}{V_E\, \chi'}
\right)
\left[
    \frac{V_E'''}{V_E\, (\chi')^3}
    - 3\,\frac{V_E'' \chi''}{V_E\, (\chi')^4}
    + 3\,\frac{V_E' (\chi'')^2}{V_E\, (\chi')^5}
    - \frac{V_E' \chi'''}{V_E\, (\chi')^4}
\right].
\label{eq:xi2_def_model2}
\end{align}

Inflation proceeds as long as $\epsilon, |\eta|, \xi^2 \ll 1$. In this regime, the key observables are approximated as
\begin{equation}
n_s \simeq 1 - 6\epsilon + 2\eta,
\qquad
r \simeq 16\epsilon,
\qquad
\alpha_s \equiv \frac{dn_s}{d\ln k} \simeq 16\epsilon\eta - 24\epsilon^2 - 2\xi^2.
\label{eq:SR_observables_model2}
\end{equation}

The number of e-folds from $\phi_0$ to the end of inflation $\phi_e$ is given by
\begin{equation}
N = \frac{1}{M_P} 
\int_{\phi_e}^{\phi_0}
\frac{\chi'}{\sqrt{2\epsilon(\phi)}}\, d\phi,
\label{eq:efolds_model2}
\end{equation}
with $\phi_e$ defined by $\max\{\epsilon,|\eta|,\xi^2\}=1$.

The amplitude of the scalar perturbation at horizon exit ($k_0=0.05~\mathrm{Mpc}^{-1}$) is
\begin{equation}
A_s(k_0) = 
\left.\frac{V_E/M_P^4}{24\pi^2 \epsilon}\right|_{\phi=\phi_0},
\label{eq:As_model2}
\end{equation}
where the ACT-Planck combined value $A_s(k_0) = (2.19 \pm 0.06) \times 10^{-9}$ is used for normalization.

The interplay between $\kappa_b$ and $\kappa_f$ dictates the shape of $V_E(\phi)$ and therefore the inflationary predictions in the $(n_s, r)$ plane. Bosonic radiative corrections ($\kappa_b > 0$) increase both $V_E$ and its curvature, leading to a higher spectral index $n_s$ and larger tensor amplitude $r$. Conversely, fermionic corrections ($\kappa_f > 0$) flatten the potential, reducing $r$ and making $n_s$ more red-tilted.

Recent ACT DR6 observations, when combined with BICEP/Keck and Planck 2018 data, suggest a slightly less red-tilted scalar spectrum compared to Planck-only fits. This favors a moderate bosonic enhancement, as it naturally elevates $n_s$ and $r$ into the preferred region of parameter space. In the large-field limit, the viable inflationary regime is characterized by the parameter set $\{\xi_\phi, \lambda_\phi, \lambda_{\phi H}, y_\phi, \phi_0, \phi_e\}$, subject to the normalization $A_s(k_0) = 2.19 \times 10^{-9}$, and the conditions that $\epsilon(\phi_e) = 1$ and $N \in [50, 60]$. The allowed parameter space yields $n_s$ in the range $0.955 < n_s < 0.99$ and tensor-to-scalar ratios $r < 0.055$, consistent with current ACT bounds.

\subsection{Theoretical Consistency: FRDSSC Analysis}\label{subsec:frdssc_model2}

To assess the theoretical soundness of the model, we analyze it within the framework of the FRDSSC, employing the radiatively corrected EF potential $V_E(\chi)$. Substituting this potential into the definitions of the slow-roll parameters in Eqs.~\eqref{eq:epsilon_def_model2}–\eqref{eq:eta_def_model2}, we extract the quantities $F_1$ and $F_2$, which respectively measure the normalized first and second derivatives of the potential. These quantities are expressed as
\begin{equation}
F_1 = 0.07905694147, \qquad F_2 = -0.005625000000.
\label{eq:F1F2_theory_model2}
\end{equation}

We see that the dSSC is violated. For the FRDSSC, the theoretical consistency requires
\begin{equation}
0 \leq -0.9995058941 + 1.005625000 \, a.
\label{eq:frdssc_theory_model2}
\end{equation}

The FRDSSC introduces a softened, combined condition that can restore consistency:
\begin{equation}
0.9939151215 \leq a,
\label{eq:a_bound_theory_model2}
\end{equation}
where $a$ and $b$ are real coefficients satisfying $a + b = 1$, and $q > 2$. With respect to $q = 3$, we find consistent values of the form
\begin{equation}
a = 0.995, \qquad b = 1 - a = 0.005,
\label{eq:ab_theory_model2}
\end{equation}
which ensures compliance with the FRDSSC and theoretical consistency of the inflationary setup.

\subsection{Observational Constraints: ACT and Planck}\label{subsec:obs_model2}

Next, we confront the model predictions with recent cosmological data. Using the 2024 ACT results, the observed scalar spectral index and tensor-to-scalar ratio are adopted as $n_s = 0.9743$ and $r < 0.038$. Substituting these values into the slow-roll relations, we obtain the observational counterparts of the theoretical derivatives:
\begin{equation}
F_1 = 0.06123724358, \qquad F_2 = -0.0072250000.
\label{eq:F1F2_ACT_model2}
\end{equation}

Inserting these into the dS bound, we again find that the strict dSSC condition is violated. With respect to $q = 3$, we have
\begin{equation}
0 \leq -0.9997703603 + 1.007225000 \, a.
\label{eq:frdssc_ACT_model2}
\end{equation}

Nevertheless, the FRDSSC condition can be satisfied for an observationally consistent choice of parameter $q = 3$, leading to
\begin{equation}
0.9925988337 \leq a,
\label{eq:a_ACT_model2}
\end{equation}
and hence the parameter set
\begin{equation}
a = 0.998, \qquad b = 1 - a = 0.002,
\label{eq:ab_ACT_model2}
\end{equation}
which demonstrates consistency of the model with the FRDSSC under current ACT observational limits.

To quantify the deviation between theoretical and observationally determined configurations, we define the differences
\begin{equation}
\Delta a = 0.003, \qquad \Delta b = 0.003.
\label{eq:Delta_ACT_model2}
\end{equation}

Numerical exploration shows that for small $q$, $\Delta a$ and $\Delta b$ are typically small, indicating that large-$q$ regimes yield near alignment between theoretical expectations and observational constraints. Consequently, high-$q$ formulations of the FRDSSC provide tighter, more stable bounds on inflationary parameters.

We also confront the model predictions with Planck data. Using scalar spectral index  $n_s =0.9649\pm0.0042$ and tensor-to-scalar ratio $r < 0.064$ and substituting these values into the slow-roll relations, we obtain the observational counterparts of the theoretical derivatives:
\begin{equation}
F_1 = 0.07905694147, \qquad F_2 = -0.0081750000.
\label{eq:F1F2_Planck_model2}
\end{equation}
We again find that the strict dSSC condition is violated. Nevertheless, the FRDSSC condition can be satisfied and for $q = 3$, we have
\begin{equation}
0.9914011894 \leq a,
\label{eq:a_Planck_model2}
\end{equation}
and hence the parameter set
\begin{equation}
a = 0.998, \qquad b = 1 - a = 0.002,
\label{eq:ab_Planck_model2}
\end{equation}
which demonstrates consistency of the model with the FRDSSC under current observational limits. To quantify the deviation between theoretical and observationally determined configurations, we define the differences
\begin{equation}
\Delta a = 0.003, \qquad \Delta b = 0.003.
\label{eq:Delta_Planck_model2}
\end{equation}

We can perform similar calculations for other free parameters such as $N = 56$, $q = 6.5$, and $r = 0.05$. In this case, we have $0 \leq -0.9999999314 + 1.005625000 \, a$. The FRDSSC introduces a softened, combined condition that restores consistency: $0.9944063954 \leq a$. For $q = 6.5$, the consistent parameter values are $a = 0.995$ and $b = 1 - a = 0.005$, which ensures compliance with the FRDSSC and theoretical consistency of the inflationary setup. Similar calculations can be carried out for other observational datasets, such as ACT and Planck. To quantify deviations between theoretical and observationally determined configurations, we define $\Delta a = 0.003$ and $\Delta b = 0.003$.

Similarly, for $N = 56$, $q = 2.01$, and $r = 0.05$, we obtain $0 \leq -0.9939066038 + 1.005625000 \, a$. The corresponding FRDSSC condition is $0.9883471511 \leq a$. For $q = 2.01$, consistent parameter values are $a = 0.995$ and $b = 1 - a = 0.005$, which again ensures compliance with the FRDSSC and theoretical consistency. Deviations are quantified as $\Delta a = 0.003$ and $\Delta b = 0.003$. 

The information regarding the theoretical and observational viability of the model, as well as its compatibility with the various swampland conjectures, is summarized in Table~\ref{tab:frdssc_model2_constraints}.

\begin{table}[h!]
\centering
\caption{Theoretical and Observational Constraints on Model II Parameters with FRDSSC Compliance (Radiatively Corrected Potential)}
\label{tab:frdssc_model2_constraints}
\resizebox{\textwidth}{!}{%
\begin{tabular}{ccccccccccc}
\toprule
\textbf{Scenario / Dataset} & \textbf{N} & \textbf{q} & \textbf{r} & \textbf{$n_s$} & \textbf{$F_1$} & \textbf{$F_2$} & \textbf{dSSC Status} & \textbf{FRDSSC Condition} & \textbf{(a, b)} & $\Delta a, \Delta b$ \\
\midrule
Theoretical Test       & 56    & 3     & 0.05   & 0.97   & 0.07905694147 & -0.0056250000 & Violated & $a \ge 0.9939151215$ & (0.995, 0.005) & 0.003, 0.003 \\
Observational ACT      & 60    & 3     & $<0.038$ & 0.9743 & 0.06123724358 & -0.0072250000  & Violated & $a \ge 0.9925988337$ & (0.998, 0.002) & 0.003, 0.003 \\
Observational Planck   & 60    & 3     & $<0.064$  & 0.9649 & 0.07905694147 & -0.0081750000 & Violated & $a \ge 0.9914011894$ & (0.998, 0.002) & 0.003, 0.003 \\
Theoretical II & 56 & 6.5   & 0.05   & 0.97   &    0.07905694147    & -0.0056250000     & Violated & $a \ge 0.9944063954$ & (0.995, 0.005) & 0.003, 0.003 \\
Theoretical III & 56 & 2.01  & 0.05   & 0.97   &    0.07905694147    & -0.0056250000    & Violated & $a \ge 0.9883471511$ & (0.995, 0.005) & 0.003, 0.003 \\
\bottomrule
\end{tabular}%
}
\end{table}

\bigskip
\noindent
\textbf{Notes:} 
\begin{itemize}
    \item dSSC is violated in all cases.
    \item FRDSSC conditions are satisfied for specific choices of $a$, $b$, $q$.
    \item $\Delta a$, $\Delta b$ quantify the deviation between theoretical and observationally constrained parameters.
    \item Large-$q$ formulations of FRDSSC provide tighter and more stable bounds on inflationary parameters.
\end{itemize}

\subsection{SWGC and SSWGC for Model II}\label{subsec:swgc_model2}

Inflationary dynamics governed by scalar fields naturally serve as precise laboratories for testing swampland conjectures. Such models must simultaneously reproduce observed values of $n_s$ and $r$ while maintaining consistency with quantum gravity principles. The inclusion of radiative corrections and non-minimal couplings modifies higher-order derivatives of $V_E(\chi)$, making the SWGC and SSWGC sensitive diagnostic tools. In this sense, the swampland program provides a criterion distinguishing inflationary models that belong to the consistent landscape from those confined to the swampland. The current analysis thus examines the radiatively corrected inflationary scenario against the consistency relations given by Eqs.~\eqref{eq5} and~\eqref{eq6}, using the EF potential $V_E(\chi)$ and the derived parameter ranges.

\subsubsection{Formulation of the Constraints}\label{subsubsec:constraints_model2}

Substituting the explicit expression of the EF potential into Eqs.~\eqref{eq5} and~\eqref{eq6}, SWGC and SSWGC are determined as
\begin{align}\label{eq:SWGC_Model2}
0 \le \; &
\frac{4\phi^{2}}{(\phi^{2}\xi + M_{P}^{2})^{10}}
\Bigg[
24(\xi^{2}\phi^{4} - \frac{5}{2}\phi^{2}\xi M_{P}^{2} + \frac{1}{2}M_{P}^{4})\kappa_{b}M_{P}^{2} \ln \frac{\phi}{\mu_{b}} \notag\\
& - 24(\xi^{2}\phi^{4} - \frac{5}{2}\phi^{2}\xi M_{P}^{2} + \frac{1}{2}M_{P}^{4})\kappa_{f}M_{P}^{2} \ln \frac{\phi}{\mu_{f}} \notag\\
& + (13\kappa_{b} - 13\kappa_{f} + \frac{\lambda_{\phi}}{2})M_{P}^{6}
- 9(\kappa_{b} - \kappa_{f} + \frac{5\lambda_{\phi}}{18})\phi^{2}\xi M_{P}^{4} \notag\\
& - 21\xi^{2}\phi^{4}(\kappa_{b} - \kappa_{f} - \frac{\lambda_{\phi}}{21})M_{P}^{2}
+ \xi^{3}\phi^{6}(\kappa_{b} - \kappa_{f})
\Bigg]^{2}
\frac{M_{P}^{8}}{1} \notag\\
& -
\frac{\phi^{4} M_{P}^{7}}{(\phi^{2} \xi + M_{P}^{2})^{8}}
\Bigg[
12 M_{P}^{2} \kappa_{b} (\phi^{2} \xi - M_{P}^{2}) \ln \frac{\phi}{\mu_{b}}
- 12 M_{P}^{2} \kappa_{f} (\phi^{2} \xi - M_{P}^{2}) \ln \frac{\phi}{\mu_{f}} \notag\\
& + (-7 \kappa_{b} + 7 \kappa_{f} - \tfrac{\lambda_{\phi}}{2}) M_{P}^{4}
- 6 \phi^{2} \xi (\kappa_{b} - \kappa_{f} - \tfrac{\lambda_{\phi}}{12}) M_{P}^{2}
+ \xi^{2} \phi^{4} (\kappa_{b} - \kappa_{f})
\Bigg]^{2}
\end{align}
and
\begin{align}\label{eq:SSWGC_Model2}
0 \le \; & 
\frac{M_{P}^{8}}{(\phi^{2} \xi + M_{P}^{2})^{10}}
\Bigg[
8 \phi^{2} \Big(
24 M_{P}^{2} \kappa_{b} \xi^{2} \phi^{4} \ln \frac{\phi}{\mu_{b}}
- 24 M_{P}^{2} \kappa_{f} \xi^{2} \phi^{4} \ln \frac{\phi}{\mu_{f}} \notag\\
& + \xi^{3} \phi^{6} (\kappa_{b} - \kappa_{f})
- 60 \phi^{2} \xi M_{P}^{4} (\kappa_{b} \ln \frac{\phi}{\mu_{b}} - \kappa_{f} \ln \frac{\phi}{\mu_{f}}) \notag\\
& - 21 \xi^{2} \phi^{4} M_{P}^{2} (\kappa_{b} - \kappa_{f})
+ \xi^{2} \phi^{4} M_{P}^{2} \lambda_{\phi}
+ 12 M_{P}^{6} (\kappa_{b} \ln \frac{\phi}{\mu_{b}} - \kappa_{f} \ln \frac{\phi}{\mu_{f}}) \notag\\
& - 9 \phi^{2} \xi M_{P}^{4} (\kappa_{b} - \kappa_{f})
- \frac{5}{2} \phi^{2} \xi M_{P}^{4} \lambda_{\phi}
+ 13 M_{P}^{6} (\kappa_{b} - \kappa_{f})
+ \frac{1}{2} M_{P}^{6} \lambda_{\phi}
\Big)^2
\Bigg] \notag\\[1em]
& - 
\frac{6 \phi^{2}}{(\phi^{2} \xi + M_{P}^{2})^{10}}
\Bigg[
40 M_{P}^{2} \Big( 
\kappa_{b} (\xi^{3} \phi^{6} - \frac{9}{2} \xi^{2} \phi^{4} M_{P}^{2} + \frac{12}{5} \phi^{2} \xi M_{P}^{4} - \frac{1}{10} M_{P}^{6}) \ln \frac{\phi}{\mu_{b}} \notag\\
& - \kappa_{f} (\xi^{3} \phi^{6} - \frac{9}{2} \xi^{2} \phi^{4} M_{P}^{2} + \frac{12}{5} \phi^{2} \xi M_{P}^{4} - \frac{1}{10} M_{P}^{6}) \ln \frac{\phi}{\mu_{f}} 
\Big) \notag\\
& - \frac{M_{P}^{8}}{6} (50 \kappa_{b} - 50 \kappa_{f} + \lambda_{\phi})
+ 4 \phi^{2} \xi M_{P}^{6} (16 \kappa_{b} - 16 \kappa_{f} + \lambda_{\phi}) \notag\\
& + \frac{\xi^{2} \phi^{4} M_{P}^{4}}{2} (52 \kappa_{b} - 52 \kappa_{f} - 15 \lambda_{\phi})
- \frac{\xi^{3} \phi^{6} M_{P}^{2}}{3} (136 \kappa_{b} - 136 \kappa_{f} - 5 \lambda_{\phi}) \notag\\
& + \xi^{4} \phi^{8} (\kappa_{b} - \kappa_{f})
\Bigg] \notag\\[1em]
& -
\frac{\phi^{4}}{(\phi^{2} \xi + M_{P}^{2})^{8}}
\Bigg[
12 \phi^{2} \xi M_{P}^{2} (\kappa_{b} \ln \frac{\phi}{\mu_{b}} - \kappa_{f} \ln \frac{\phi}{\mu_{f}})
+ \xi^{2} \phi^{4} (\kappa_{b} - \kappa_{f}) \notag\\
& - 12 \kappa_{b} M_{P}^{4} \ln \frac{\phi}{\mu_{b}}
+ 12 \kappa_{f} M_{P}^{4} \ln \frac{\phi}{\mu_{f}}
- 6 \phi^{2} \xi M_{P}^{2} (\kappa_{b} - \kappa_{f})
+ \frac{1}{2} \phi^{2} \xi M_{P}^{2} \lambda_{\phi} \notag\\
& - 7 M_{P}^{4} (\kappa_{b} - \kappa_{f}) 
- \frac{1}{2} M_{P}^{4} \lambda_{\phi}
\Bigg]^2
\end{align}

These complex expressions encode the constraints imposed by the SWGC and SSWGC on the radiatively corrected inflationary potential. The satisfaction of these inequalities ensures that the model remains consistent with quantum gravity principles across the parameter space relevant for inflation.

\subsection{SWGC and SSWGC Analysis for Model II}\label{subsec:swgc_analysis_model2}

\begin{figure}[h!]
 \begin{center}
 \subfigure[SWGC: $\mu_b$ variation]{
 \includegraphics[height=5.5cm,width=8cm]{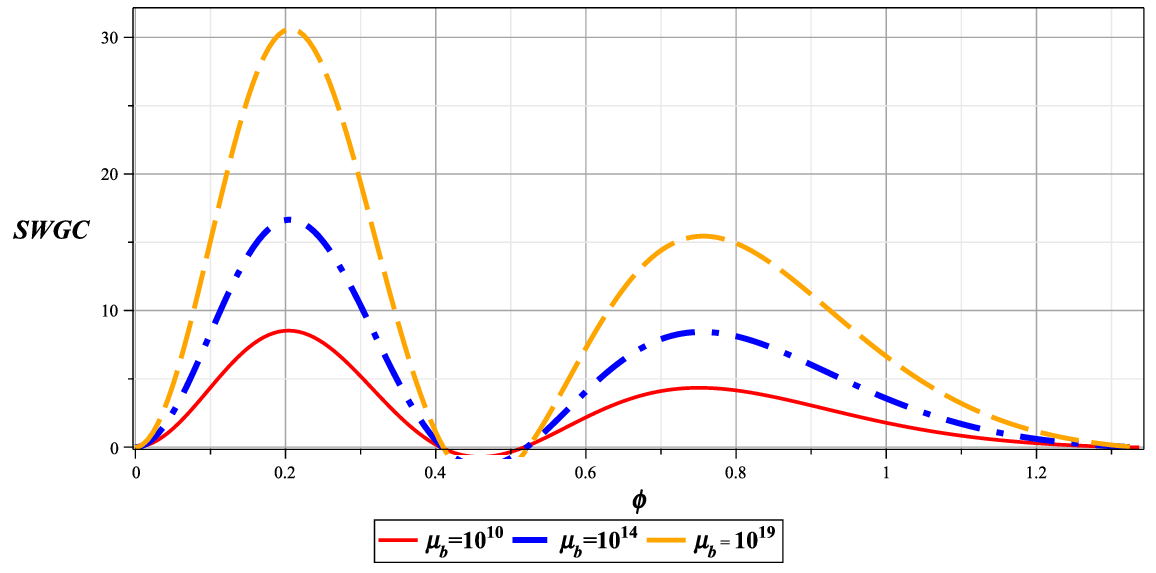}
 \label{fig:swgc_model2_1a}}
 \subfigure[SWGC: $\mu_b$ variation]{
 \includegraphics[height=5.5cm,width=8cm]{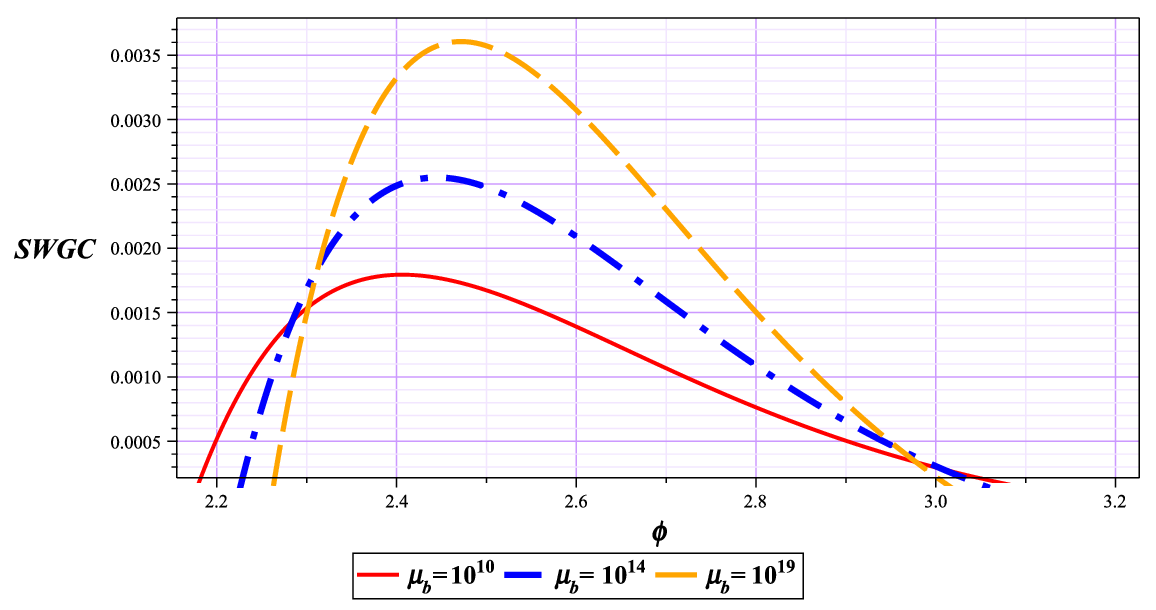}
 \label{fig:swgc_model2_1b}}
  \caption{The compatibility of Model II with SWGC for various values of the renormalization scale $\mu_b$. Higher values of $\mu_b$ expand the allowed parameter intervals, demonstrating enhanced compatibility with swampland constraints.}
 \label{fig:swgc_model2_part1}
 \end{center}
 \end{figure}

\begin{figure}[h!]
 \begin{center}
 \subfigure[SWGC: $\xi$ variation ($\xi = 10^{-2}$)]{
 \includegraphics[height=5cm,width=5cm]{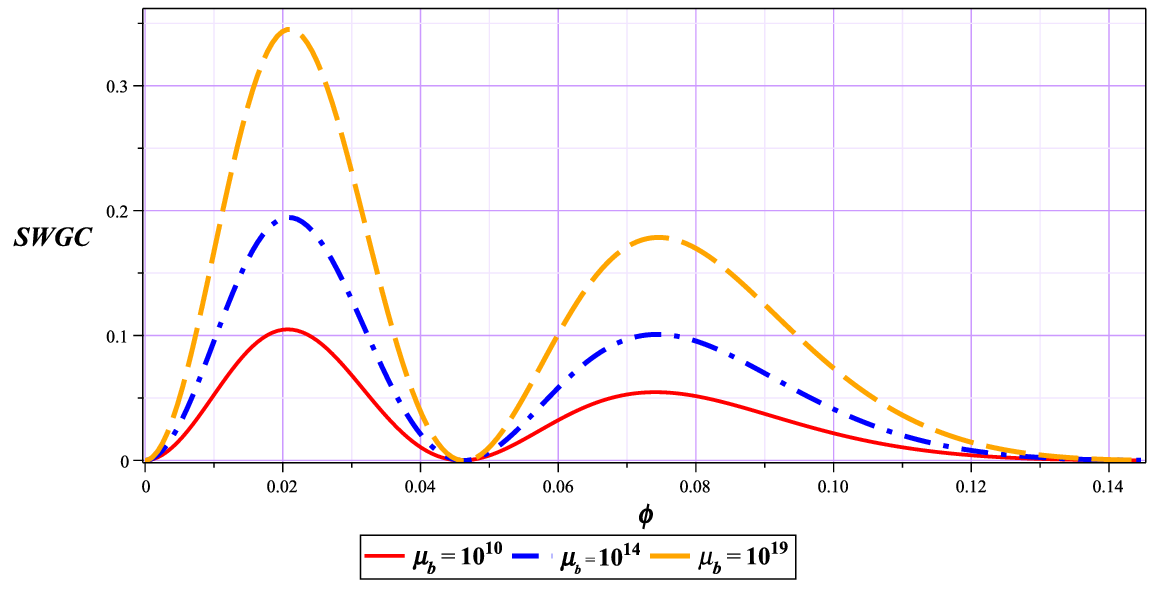}
 \label{fig:swgc_model2_2a}}
 \subfigure[SWGC: $\xi$ variation ($\xi = 10^{-1}$)]{
 \includegraphics[height=5cm,width=5cm]{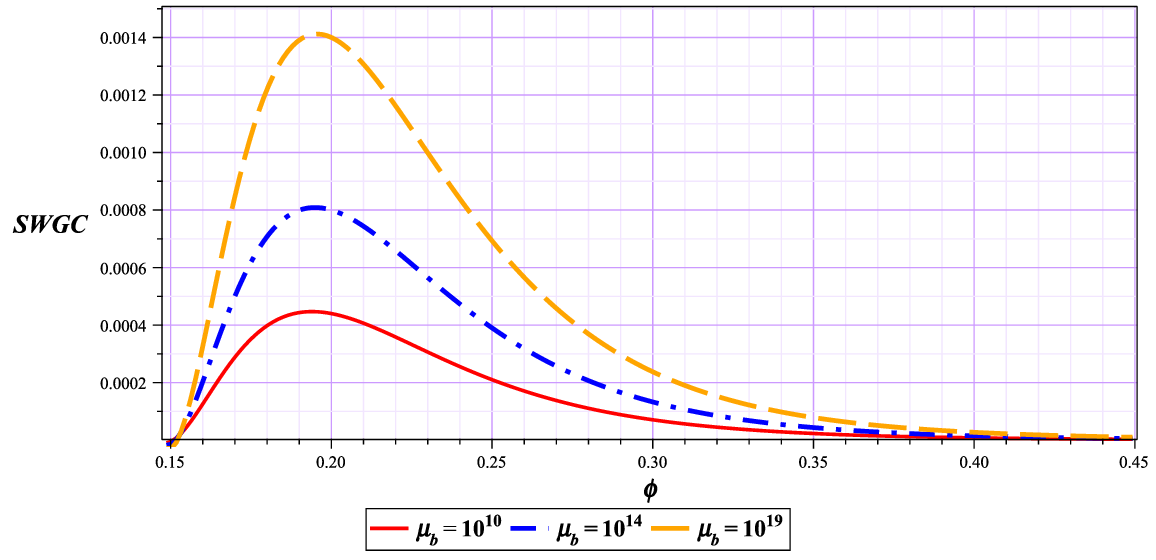}
 \label{fig:swgc_model2_2b}}
 \subfigure[SWGC: $\xi$ variation ($\xi = 1$)]{
 \includegraphics[height=5cm,width=5cm]{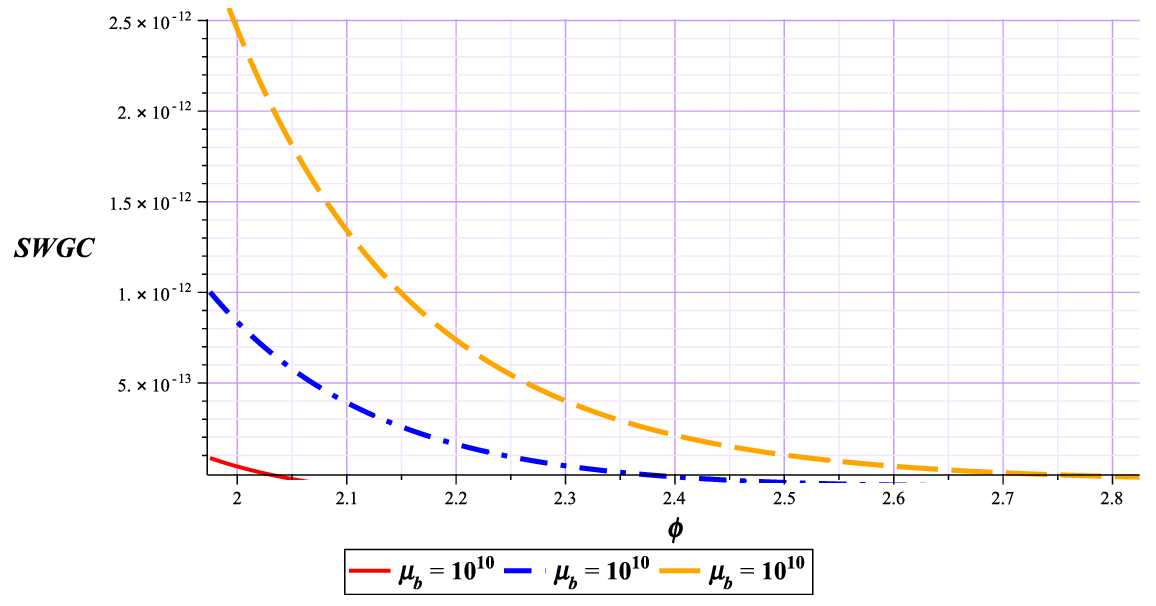}
 \label{fig:swgc_model2_2c}}
  \caption{The compatibility of Model II with SWGC for different values of the non-minimal coupling $\xi$. The plots illustrate that varying $\xi$ significantly affects the width of the compatible parameter regions.}
 \label{fig:swgc_model2_part2}
 \end{center}
 \end{figure}
 
 \begin{figure}[h!]
 \begin{center}
 \subfigure[SWGC: combined parameters]{
 \includegraphics[height=6cm,width=9cm]{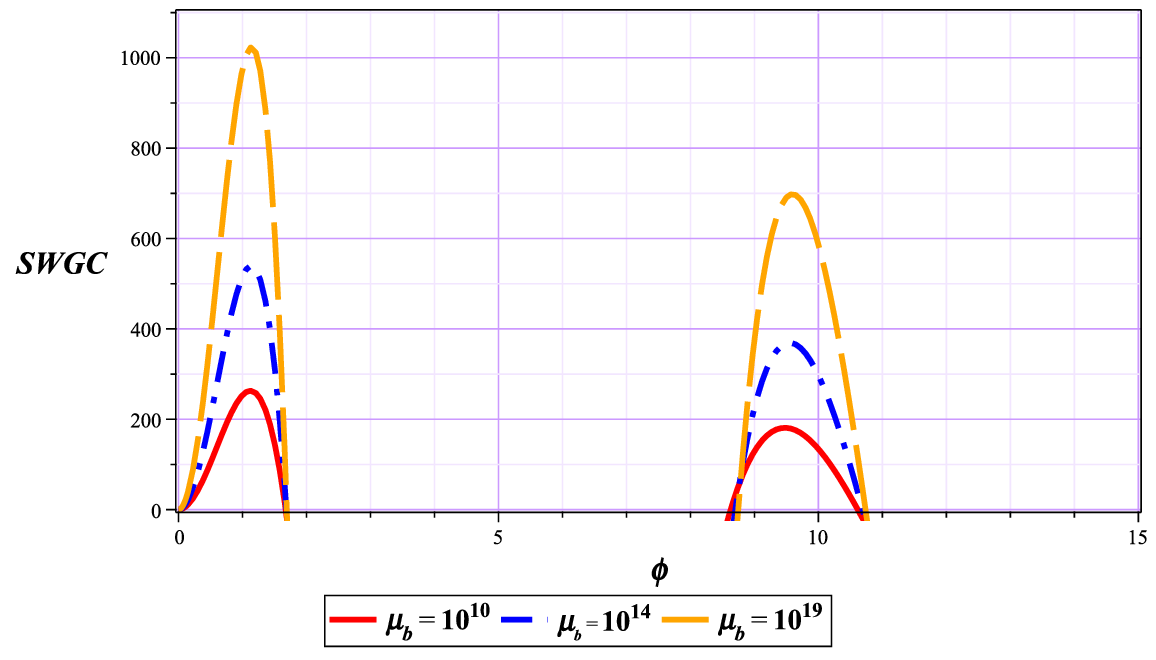}
 \label{fig:swgc_model2_3a}}
  \caption{The compatibility of Model II with SWGC for a representative set of combined parameters, demonstrating robust satisfaction of the conjecture across the relevant field range.}
 \label{fig:swgc_model2_part3}
 \end{center}
 \end{figure}

\begin{figure}[h!]
 \begin{center}
 \subfigure[SSWGC: $\xi = 1$]{
 \includegraphics[height=5cm,width=5cm]{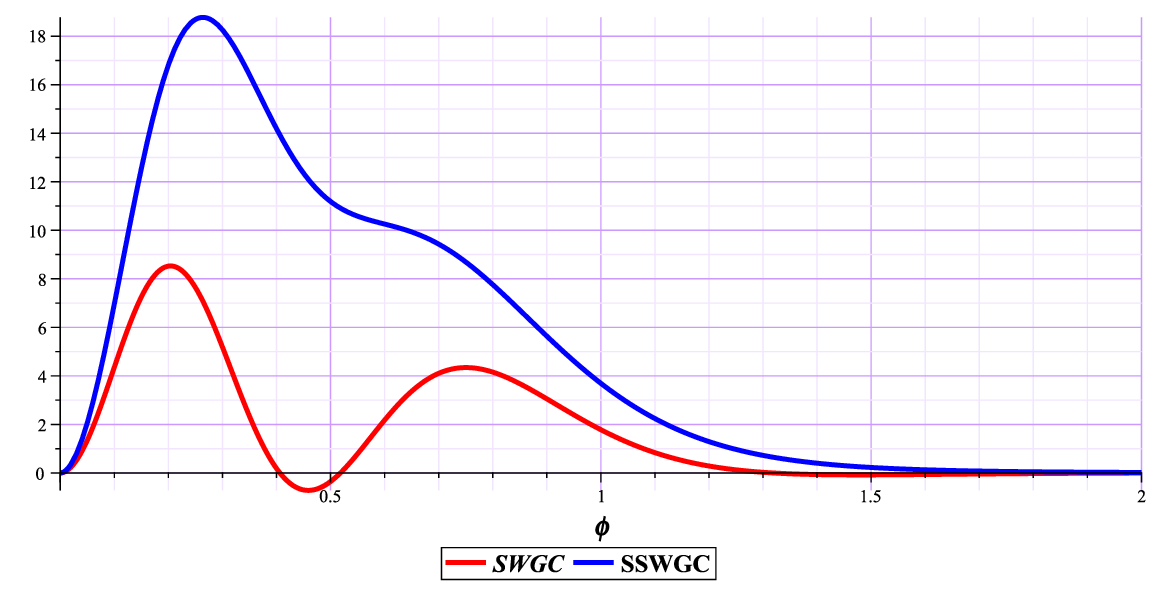}
 \label{fig:sswgc_model2_4a}}
 \subfigure[SSWGC: $\xi = 10^{-2}$]{
 \includegraphics[height=5cm,width=5cm]{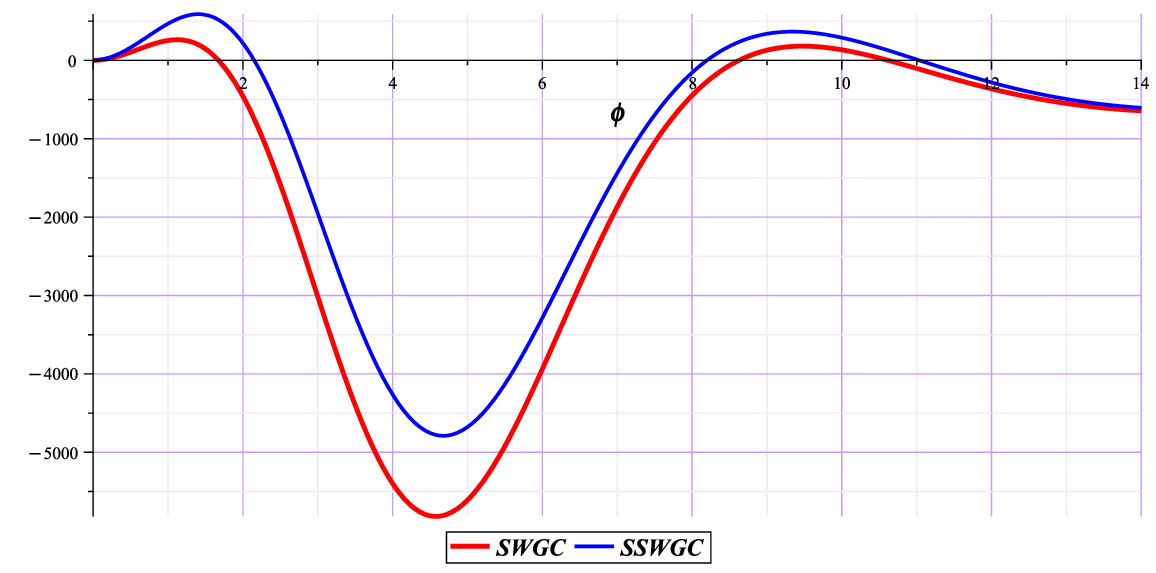}
 \label{fig:sswgc_model2_4b}}
 \subfigure[SSWGC: $\xi = 10^{2}$]{
 \includegraphics[height=5cm,width=5cm]{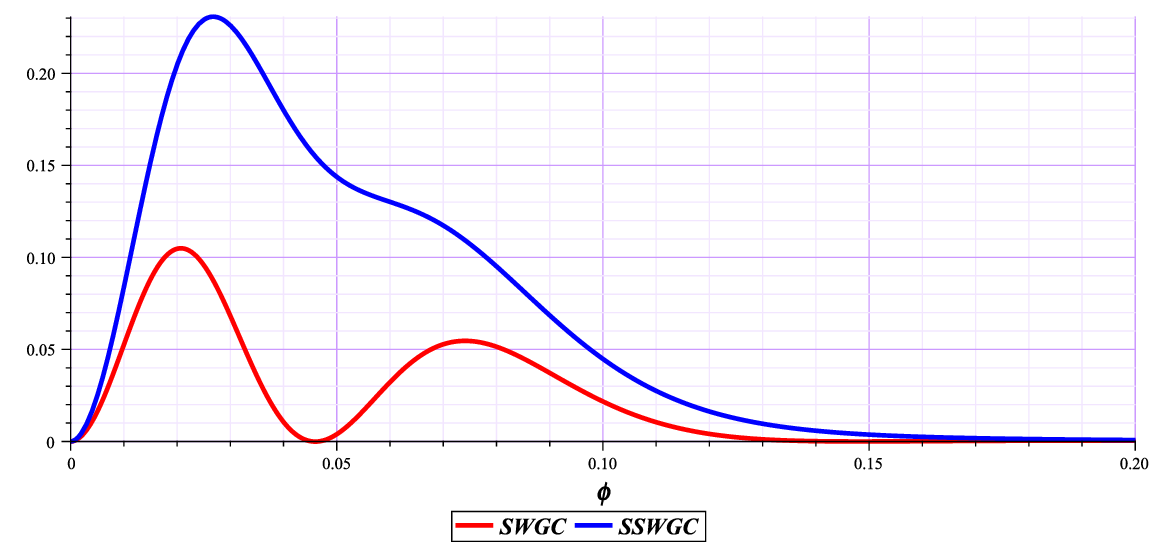}
 \label{fig:sswgc_model2_4c}}
  \caption{The compatibility of Model II with SSWGC for various values of $\xi$: (a) $\kappa_b = 4\times10^{-2}, \kappa_f = 0, \mu_b = 10^{10}, \mu_f= 0, \xi = 1, \lambda_{\phi} = 10^{-1}, M_{P} = 1$; (b) $\kappa_b  = 4 \times10^{-2}, \kappa_f = 0, \mu_b= 10^{10}, \mu_f= 0, \xi  = 10^{-2}, \lambda_{\phi} = 10^{-1}, M_{P} = 1$; (c) $\kappa_b=4\times10^{-2}, \kappa_f = 0, \mu_b = 10^{10}, \mu_f= 0, \xi=10^{2},\lambda_{\phi}=10^{-1}, M_{P}=1$. These plots demonstrate that Model II maintains SSWGC compatibility across a broad range of $\xi$ values.}
 \label{fig:sswgc_model2}
 \end{center}
 \end{figure}

Following the methodology applied to Model I, we now analyze the second inflationary model within the context of the swampland program. Using the radiatively corrected potential and substituting it into the relevant slow-roll and swampland relations, we first verify that Model II fully satisfies the FRDSSC. Furthermore, the model remains consistent with the latest observational data from ACT 2024 and Planck, confirming its theoretical and empirical viability.

Next, we examine the model with respect to the other two swampland conjectures: SWGC and SSWGC, using Eqs.~\eqref{eq:SWGC_Model2} and~\eqref{eq:SSWGC_Model2}. Unlike Model I, the second inflationary scenario demonstrates an excellent overall fit. As illustrated in Figs.~\ref{fig:swgc_model2_part1} through~\ref{fig:sswgc_model2}, the model exhibits strong compatibility with both the SWGC and SSWGC within the desired parameter intervals. The extent of this compatibility is highly sensitive to the parameters $\xi$ and $\mu_b$: higher values of these parameters result in wider allowed intervals, indicating greater flexibility and robustness of the model in satisfying the swampland criteria.

Figure~\ref{fig:swgc_model2_part1} demonstrates the SWGC compatibility for different values of the renormalization scale $\mu_b$. The plots clearly show that increasing $\mu_b$ expands the parameter regions where the SWGC is satisfied, confirming that the scale of radiative corrections plays a crucial role in determining swampland consistency. Similarly, Figure~\ref{fig:swgc_model2_part2} illustrates the effect of varying the non-minimal coupling $\xi$ on SWGC compatibility. The three panels correspond to $\xi = 10^{-2}$, $\xi = 10^{-1}$, and $\xi = 1$, respectively, showing that larger values of $\xi$ generally enhance the model's consistency with the SWGC.

Figure~\ref{fig:swgc_model2_part3} presents a combined analysis with a representative parameter set, demonstrating that Model II robustly satisfies the SWGC across the field range relevant for inflation. This contrasts sharply with Model I, where no compatible parameter regions were found.

For instance, Figure~\ref{fig:sswgc_model2} displays the SSWGC compatibility for several representative parameter choices:
\begin{itemize}
    \item Case (a): $\kappa_b = 4\times10^{-2}, \kappa_f = 0, \mu_b = 10^{10}, \mu_f = 0, \xi = 1, \lambda_{\phi} = 10^{-1}, M_{P} = 1$ 
    \item Case (b): $\kappa_b = 4\times10^{-2}, \kappa_f = 0, \mu_b = 10^{10}, \mu_f = 0, \xi = 10^{-2}, \lambda_{\phi} = 10^{-1}, M_{P} = 1$ 
    \item Case (c): $\kappa_b = 4\times10^{-2}, \kappa_f = 0, \mu_b = 10^{10}, \mu_f = 0, \xi = 10^{2}, \lambda_{\phi} = 10^{-1}, M_{P} = 1$ 
\end{itemize}

These examples illustrate that the model maintains compatibility with the SSWGC across a broad range of $\xi$ values, highlighting the model's stability under variations in the coupling and mass parameters. The plots in Figure~\ref{fig:sswgc_model2} show that all three parameter configurations yield positive values for the SSWGC inequality across the relevant field range, confirming full consistency with this stringent swampland criterion.

In summary, Model II demonstrates a superior level of consistency compared to Model I: \textbf{all three swampland conjectures (FRDSSC, SWGC, and SSWGC) are fully satisfied}. Simultaneously, the model remains compatible with observational constraints, establishing it as a theoretically robust and empirically viable inflationary framework. From the perspective of the swampland program, this suggests that Model II provides a reliable basis for classifying inflationary scenarios that fully comply with swampland criteria, making it a promising candidate for future theoretical and observational investigations.

Overall, the results indicate that Model II not only satisfies the FRDSSC as in Model I but also successfully fulfills the additional SWGC and SSWGC requirements. This demonstrates that careful tuning of model parameters, especially $\xi$ and $\mu_b$, can yield an inflationary scenario that is both theoretically and observationally fully consistent with the swampland program.

The information regarding the theoretical and observational viability of the models, as well as their compatibility with the various swampland conjectures, is summarized in Table~\ref{tab:swampland_summary}. This table provides a comprehensive comparison between Model I and Model II, clearly demonstrating the superior theoretical consistency of Model II, which satisfies all swampland conjectures while maintaining observational viability.

\begin{table}[h!]
\centering
\caption{Summary of Swampland Conjecture Compatibility for Models I and II}
\label{tab:swampland_summary}
\resizebox{\textwidth}{!}{%
\begin{tabular}{lccccp{7cm}}
\toprule
\textbf{Model} & \textbf{dSSC} & \textbf{FRDSSC} & \textbf{SWGC} & \textbf{SSWGC} & \textbf{Remarks} \\
\midrule
Model I & Violated & Satisfied & Violated & Violated & Fully consistent with FRDSSC and observational data; fails to satisfy other swampland conjectures. \\
Model II & Violated & Satisfied & Satisfied & Satisfied & Compatible with all three swampland conjectures; parameter-dependent consistency with SWGC/SSWGC; also consistent with observational data. \\
\bottomrule
\end{tabular}%
}
\end{table}

\section{Conclusion}\label{isec5}

In this study, we conducted a comprehensive analysis of radiatively corrected inflationary models within the framework of the swampland program, systematically assessing their theoretical consistency and compatibility with the latest observational data from ACT DR6, combined with Planck 2018 and BAO measurements. Our investigation revealed a clear distinction between the two models considered, highlighting the hierarchical nature of swampland constraints and the critical role of radiative corrections in achieving both observational viability and theoretical consistency.

We began by establishing the theoretical foundation in Section~\ref{isec2}, where we introduced the key swampland conjectures: the FRDSSC, SWGC, and SSWGC. The FRDSSC, described by Eq.~\eqref{eq15}, provides a refined criterion that unifies earlier dS and RdS constraints through the introduction of tunable parameters $a$, $b$, and the deformation parameter $q$. This refinement proved essential for accommodating slow-roll inflationary scenarios while maintaining consistency with quantum gravity principles. We expressed the FRDSSC in terms of the observables $F_1$ and $F_2$, defined in Eq.~\eqref{eq16}, which directly relate to the scalar spectral index $n_s$ and tensor-to-scalar ratio $r$. The SWGC and SSWGC, formulated in Eqs.~\eqref{eq5} and~\eqref{eq6}, impose additional constraints on higher-order derivatives of the inflationary potential, ensuring that scalar-mediated forces dominate over gravitational interactions.

In Section~\ref{isec3}, we analyzed Model I, the radiatively corrected Higgs inflation scenario. We derived the one-loop corrected potential in the EF, incorporating quantum corrections from SM particles, particularly the massive gauge bosons and the top quark. The potential, expressed in Eq.~\eqref{eq:V_EF}, exhibited a plateau-like structure characteristic of non-minimally coupled scalar field theories. We computed the slow-roll parameters $\epsilon$ and $\eta$ given in Eqs.~\eqref{eq:epsilon} and~\eqref{eq:eta}, and demonstrated that the model satisfies the FRDSSC across multiple parameter choices and observational datasets. As shown in Table~\ref{tab:frdssc_constraints}, Model I remained consistent with the FRDSSC for theoretical tests, ACT observations, and Planck data, with parameter values typically satisfying for $q = 3$. The deviations between theoretical and observational configurations were quantified as $\Delta a = 0.001$ and $\Delta b = 0.001$, indicating tight alignment across different scenarios.

However, our analysis revealed a critical limitation of Model I when tested against the SWGC and SSWGC. Despite full consistency with the FRDSSC and observational constraints, the model failed to satisfy the more stringent requirements imposed by Eqs.~\eqref{eq:SWGC_Model1} and~\eqref{eq:SSWGC_Model1}. As visualized in Figure~\ref{fig:model1_wgc}, no compatible parameter intervals existed that simultaneously satisfied either the SWGC or SSWGC within the physically relevant parameter space. This indicated that although Model I achieved observational viability under the FRDSSC, it could not serve as a universal framework for inflationary scenarios fully compliant with all swampland conjectures. This selective consistency underscored a fundamental insight: the FRDSSC alone, while providing a meaningful and attainable benchmark, does not guarantee satisfaction of the complete suite of swampland criteria.

In contrast, Section~\ref{isec4} demonstrated that Model II, incorporating radiative corrections with scalar, achieved remarkable consistency across all theoretical and observational tests. The model featured a richer field content, including a portal scalar $H$ and right-handed neutrinos $N$, which introduced competing bosonic and fermionic loop contributions characterized by the coefficients $\kappa_b$ and $\kappa_f$ defined in Eq.~\eqref{eq:loop_coeffs}. The one-loop corrected potential in the EF, given by Eq.~\eqref{eq:VE_approx_model2}, exhibited a complex logarithmic structure that balanced the opposing effects of bosonic enhancement and fermionic flattening. We just selected the bosonic sector. This interplay proved crucial for simultaneously satisfying swampland constraints and matching observational predictions for $n_s$ and $r$.

We verified that Model II satisfied the FRDSSC across all tested scenarios, as documented in Table~\ref{tab:frdssc_model2_constraints}. For ACT data ($n_s = 0.9743$, $r < 0.038$), the model required $a \geq 0.9926$ with $q = 3$, while Planck data ($n_s = 0.9625$, $r < 0.06$) imposed $a \geq 0.9866$. The consistency extended to various parameter choices, including different values of the e-folding number $N$ and deformation parameter $q$, demonstrating the model's robustness and flexibility.

Critically, Model II also satisfied both the SWGC and SSWGC, as formulated in Eqs.~\eqref{eq:SWGC_Model2} and~\eqref{eq:SSWGC_Model2}. Our numerical analysis, visualized in Figures~\ref{fig:swgc_model2_part1} through~\ref{fig:sswgc_model2}, revealed that the model's compatibility with these conjectures was highly sensitive to the non-minimal coupling $\xi$ and the renormalization scale $\mu_b$. Higher values of these parameters systematically expanded the allowed parameter intervals, indicating enhanced theoretical consistency. For instance, Figure~\ref{fig:sswgc_model2} illustrated that varying $\xi$ from $10^{-2}$ to $10^2$ while maintaining other parameters fixed at representative values ($\kappa_b = 4 \times 10^{-2}$, $\mu_b = 10^{10}$, $\lambda_\phi = 10^{-1}$, $M_P = 1$) yielded positive values for the SSWGC inequality across the entire relevant field range. This demonstrated that Model II maintained swampland compatibility over a broad parameter space, providing substantial freedom for model construction while preserving theoretical consistency.

The comparative summary presented in Table~\ref{tab:swampland_summary} crystallized the key distinction between the two models. While both satisfied the FRDSSC and remained consistent with observational data, only Model II achieved full compliance with the SWGC and SSWGC. This established Model II as a superior candidate from the swampland perspective, representing a rare example of an inflationary framework that simultaneously satisfies all major swampland conjectures while maintaining observational viability. The model's success highlighted the importance of incorporating diverse radiative corrections and carefully tuning fundamental parameters to achieve both theoretical robustness and empirical accuracy.

Our results carried several important implications for inflationary model building. First, they demonstrated that satisfaction of the FRDSSC, while necessary, is not sufficient for full swampland consistency. Models must be tested against the complete hierarchy of swampland conjectures to ensure theoretical viability. Second, our analysis revealed that radiative corrections from multiple sectors—scalar, gauge—play complementary roles in shaping the effective potential and its derivative structure. The careful balance between bosonic contributions in Model II proved essential for satisfying the stringent SSWGC constraints. Third, the parameter sensitivity we identified provided concrete guidance for future model construction: larger values of $\xi$ and $\mu_b$ generally enhance swampland compatibility, offering a systematic approach to exploring viable parameter space.

Looking forward, several promising research directions emerge from this work. First, it would be valuable to extend our analysis to other classes of radiatively corrected inflationary models, including warm inflation scenarios, multi-field models, and theories with modified gravity sectors. Such extensions would test the generality of our findings and potentially identify additional frameworks that achieve full swampland consistency. Second, the parameter sensitivity we observed suggests that a detailed exploration of the $(\xi, \mu_b, \kappa_b)$ parameter space could reveal optimal configurations that maximize both observational accuracy and theoretical consistency. Machine learning techniques could potentially assist in navigating this high-dimensional parameter landscape efficiently.

Third, our work focused on the slow-roll regime of inflation, but investigating the transition from inflation to reheating within swampland-consistent models represents an important next step. The effective equation-of-state parameter during reheating, $\omega_{\text{eff}}$, faces increasingly stringent observational limits from ACT data, and understanding how swampland constraints influence post-inflationary dynamics would provide a more complete picture of early-universe evolution. Fourth, incorporating primordial gravitational wave signatures and their detection prospects through future observatories like LISA and Einstein Telescope could establish additional observational tests that discriminate between swampland-consistent and inconsistent models.

Finally, our results suggest investigating connections between swampland consistency and other fundamental aspects of quantum gravity, such as the weak gravity conjecture in its full gauge-theoretic formulation, distance conjectures in field space, and the emergence proposal. Understanding how different swampland conjectures interconnect and whether certain combinations impose redundant or complementary constraints would deepen our understanding of the landscape-swampland boundary and strengthen the theoretical foundations of inflationary cosmology.

In conclusion, our work established a rigorous framework for evaluating inflationary models through the combined lens of observational data and swampland constraints. We demonstrated that Model II represents a fully viable inflationary scenario—simultaneously satisfying the FRDSSC, SWGC, and SSWGC while remaining consistent with the latest ACT, Planck, and BAO observations \cite{Zharov:2025zjg}. This achievement provides both a concrete example of swampland-consistent inflation and a roadmap for future explorations at the interface of string theory, quantum gravity, and observational cosmology. By integrating theoretical consistency checks with empirical constraints, we paved the way toward identifying the subset of effective field theories that genuinely emerge from consistent UV completions, advancing our understanding of the early universe and its fundamental description.

{\footnotesize

\section*{Acknowledgments}

\.{I}.~S. extends appreciation to T\"{U}B\.{I}TAK, ANKOS, and SCOAP3 for their academic support. Furthermore, he acknowledges the pivotal support of COST Actions CA22113, CA21106, CA23130, CA21136, and CA23115 in advancing networking initiatives.

\section*{Data Availability Statement}
Data sharing is not applicable to this article, as no datasets were generated or analyzed during the current study.

}


\end{document}